\documentclass[pdflatex,sn-mathphys]{sn-jnl}

\usepackage{Abv}
\usepackage{physics}
\usepackage[T1]{fontenc}
\usepackage[resetlabels,labeled]{multibib}
\newcounter{suppfigure}
\renewcommand{\thesuppfigure}{\arabic{suppfigure}}

\newenvironment{suppfigure}{
    \renewcommand{\figurename}{Supplementary Figure}%
    \setcounter{suppfigure}{\value{figure}} 
    \stepcounter{suppfigure}%
    \begin{figure}
}{
    \end{figure}
    \setcounter{figure}{\thesuppfigure} 
}
\usepackage[symbol]{footmisc}

\newcites{Sup}{Supplemental References}

\usepackage{xcolor}
\definecolor{mod}{rgb}{0,0,0}
\definecolor{mod_ver2}{rgb}{0,0,0}
\definecolor{mod_ver3}{rgb}{0,0,0}
\newcommand{\fix}[1]{\textcolor{mod_ver3}{#1}}

\usepackage{xr-hyper}

\begin{document}

\title[Origin of compact exoplanetary systems]{Origin of compact exoplanetary systems during disk infall}


\author*[1]{\fnm{Raluca} \sur{Rufu$^\dag$}}\email{raluca@boulder.swri.edu}

\author[1]{\fnm{Robin M.} \sur{Canup}}

\affil[1]{\orgdiv{Solar System Science and Exploration Division}, \orgname{Southwest Research Institute}, \orgaddress{\street{1301 Walnut St, Ste 400}, \city{Boulder}, \postcode{80302}, \state{CO}, \country{USA}}}

\abstract{Exoplanetary systems that contain multiple planets on short-period orbits appear to be prevalent in the current observed exoplanetary population, yet the processes that give rise to such configurations remain poorly understood. A common prior assumption is that planetary accretion commences after the infall of gas and solids to the circumstellar disk ended. However, observational evidence indicates that accretion may begin earlier. We propose that compact systems are surviving remnants of planet accretion that occurred during the final phases of infall. In regions of the disk experiencing ongoing infall, the planetary mass is set by the balance between accretion of infalling solids and the increasingly rapid inward migration driven by the surrounding gas as the planet grows. This balance selects for similarly-sized planets whose mass is a function of infall and disk conditions. We show that infall-produced planets can survive until the gas disk disperses and migration ends, and that across a broad range of conditions, the mass of surviving systems is regulated to a few $\times 10^{-5}$ to $10^{-4}$ times the host star's mass. This provides an explanation for the similar mass ratios of known compact systems.}

\maketitle

\footnotetext[\value{footnote}]{\textsuperscript{\dag} NASA Hubble Fellowship Program Sagan Fellow}
\section*{Introduction}
\begin{figure}
       \centering
      \includegraphics[width=\textwidth]{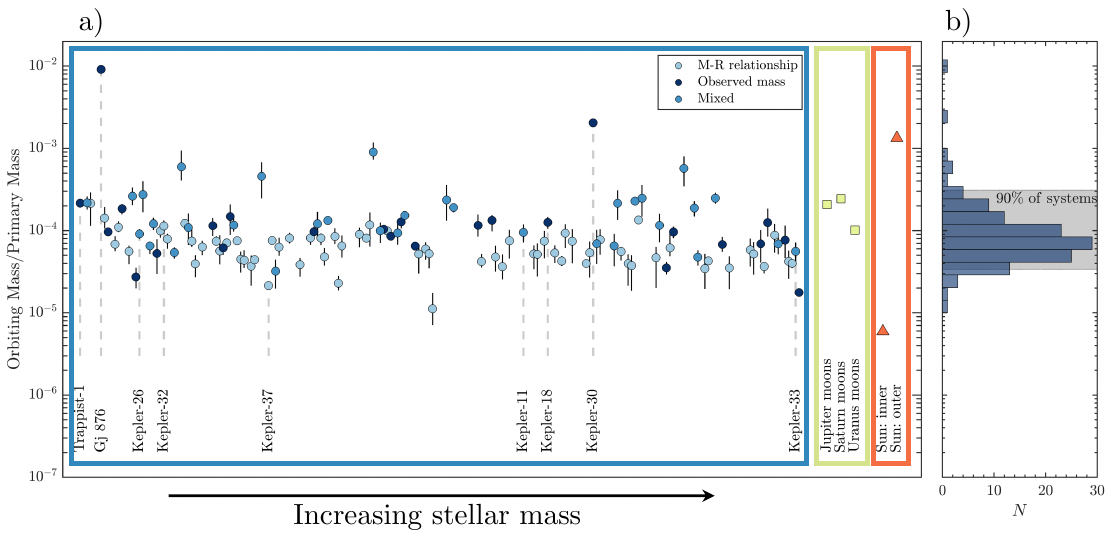}
      \caption{\textbf{Estimated total mass of transiting compact  systems, $M_{\rm tot}$, scaled to the stellar mass, $M_*$}.  \textbf{a,} $(M_{\rm tot}/M_*)$ for compact systems having $\geq 3$ known planets that orbit a single star within $a<0.5$ au (circle markers, blue box).  Points are ordered left-to-right by ascending stellar mass. Data are from
      \textit{www.exoplanetarchive.ipac.caltech.edu}. For cases without mass estimates, we use the observed planet radius, increase the estimated radius uncertainty by a factor of 2, and then apply a radius vs. mass  relation \cite{WeissMarcy2014}. Light [medium] blue circles are systems with all [some] planetary masses estimated from this relation, while dark blue circles are systems with measured planetary masses. We include only systems with resulting errors  $\Delta M_{\rm tot}/M_{\rm tot}<1$.  \textbf{b,} Distribution of compact system mass ratios. Over a wide range of stellar masses ($M_* \sim 0.1$ to $1.3$ stellar mass), compact multi-planet systems display a common mass ratio, with $90\%$ of systems having $3\times 10^{-5} < (M_{\rm tot}/M_*) < 3\times 10^{-4}$ (gray shaded region).  This mass ratio is more similar to that of the gas giant satellite systems (square markers, yellow box) than to the inner or outer planets in our Solar System (triangle markers, red box). }
      \label{fig:MassRatio}
\end{figure}

Compact exoplanetary systems have multiple planets that are mostly solid by mass, with approximately co-planar and circular orbits at small radial distances of only about 10 to $10^2$ times their host star's radius($R_*$)\cite{weiss2022architectures}.  Compact systems also display a similar ratio of the total mass of planets in each system compared to the stellar mass (blue box, Figure \ref{fig:MassRatio}). While there could be undetected systems with smaller mass ratios, the largest close-in planets have likely been detected, and so the maximum mass ratio of approximately $10^{-4}$ appears significant. This is not explained solely by dynamical stability: e.g., a Trappist-1-like star could accommodate three $5$ Earth-mass planets totaling  about $6 \times 10^{-4}$ times the stellar mass, $M_*$, but such large mass ratio systems seem rare.  

How compact systems formed and why they are so different than our Solar System is a fundamental question. A  puzzle is how planets with orbital periods of only days-to-weeks survived within a circumstellar gas disk that persisted for about $10^6$ yr. Gravitational interactions with a gas disk typically cause a planet's orbit to spiral inward, which would lead to rapid planet loss in a compact disk with a stellar metallicity.  For some disk conditions, orbital migration is outward \cite{PaardekooperMellema2006}, but these must not have predominated in compact systems whose planets remained close to the star. Prior works have shown that planet loss may be prevented if the gas disk has an inner cavity that halts migration of planets near its outer edge \citep{OgiharaIda2009, Ormel2017,Izidoro2017,Izidoro2021,batygin2023formation}. Such cavities are common around T Tauri (Class II) stars \cite{zhu2023global}.


However, there has been no explanation for the common mass ratio among compact systems seen in Figure \ref{fig:MassRatio}.  Prior models presume a priori a starting mass of accreting disk solids equal to that needed to yield the observed compact system planet masses. Yet what established that initial mass $--$ and why resulting compact systems would display a preferred mass ratio despite varied stellar masses and disk evolutions $--$ has been a mystery. If the preferred ratio were simply due to some process having selected for a common disk-to-star mass ratio, then the mass of solids available for forming compact planets would be proportional to the stellar metallicity.  Instead, the occurrence of compact systems shows a  weak dependence on metallicity, in contrast to the strong metallicity dependence of hot/cold Jupiters \citep{petigura2018california}. These issues motivate additional consideration of the processes that establish the initial conditions for compact system formation. 

As a molecular cloud core collapses, a circumstellar disk forms because infalling gas and entrained grains have too much angular momentum to fall directly onto the star. 
The timescale over which infall occurs is constrained by surveys of young stellar objects.  Estimated median lifetimes (half-lives), $t_{1/2}$, for Class 0 + I objects undergoing infall range from about $0.1$ to 0.5 Myr, with an additional $0.1$ to 0.4 Myr 
spent in a subsequent ``flat SED" phase \cite{dunham2014evolution,kristensen2018protostellar}, which may be associated with remnant infall \cite{dunham2014evolution,Mottram2017}. 

A nearly universal assumption is that planet accretion begins only after infall ends. Yet, increasing observational evidence points to an earlier start to accretion. 
The mass in sub-cm particles in 1 to 3 Myr old disks seems too low to explain observed compact systems, suggesting that most solids have accreted into larger sizes by this time \cite{Manara2018,DashMiguel2020}. The flux density at different wavelengths in systems undergoing infall is indicative of grain growth \cite{tychoniec2020dust}, and there is evidence for mm-sized pebbles in the collapsing envelopes and disks of protostars that are only about $10^5$ yr old  \cite{Miotello2014,Harsono2018}. Further, large-scale dust rings and gaps in disks undergoing infall hint that 
large bodies may exist or be forming by this time \cite{SeguraCox2020, Alves2020}.

In this work we develop a concept in which compact system planets form during the end stages of infall.  Planets accrete close to the star with their masses regulated by a balance between mass accumulation from the infalling solids and the inward gas-driven migration that becomes faster as the planetary mass increases. Across a wide range of disk conditions, we find that this balance naturally leads to total planetary system masses of a few × $10^{-5}$ to $10^{-4}$ times the stellar mass, matching the observed properties of compact systems (Figure \ref{fig:MassRatio}).

\section*{Results}
We adopt a standard model in which gas and tiny grains infall to a circumstellar disk interior to the centrifugal radius, $r_c$ (Supplemental Figure \ref{fig:Schematic}), with the gas then spreading viscously
\cite{HuesoGuillot2005}. The detailed properties of the collapsing core would determine $r_c$ 
\cite{HuesoGuillot2005,Hennebelle2016} 
(Supplementary Figure \ref{fig:Schematic}). Observed disks undergoing infall have mean radii of about a few $\times 10$ au, with wide scatter down to observational limits at approximately $10$ au \cite{Tobin2020}.  
Because some disk material viscously expands beyond $r_c$ during infall, observed disk radii would be larger than $r_c$. 
We here consider that compact systems result from disks with small $r_c \le 1$ au, much less than the $10$ au often assumed for our solar system.

\textcolor{mod}{Once in stellar orbit, grains delivered to the disk by the infall undergo mutual collisions, particle growth, and gas-driven radial drift \cite{birnstiel2010gas}.  
How grains grow into km-sized planetesimals and larger is a complex question  \cite{birnstiel2023dust}. A premise of our model is that planetesimals \fix{form somewhere within the region of infall} (i.e., at or interior to $r_c$) during the final infall stage.} 
Planetesimal formation in inner disks is plausible given recent results (see also \nameref{sec:PlanetesimalAcc} section in Methods). Experiments find a 2 to 3 orders-of-magnitude increase in rocky grain adhesive forces at high inner disk temperatures 
\cite{pillich2023composition}, and observations indicate particle growth in young disks interior to the ice line \cite{liu2021millimeter,yamamoto2014examination,zamponi2024exploring}, suggesting 
that cm-scale and larger pebbles may form interior to $r_c$. Planetesimal formation via streaming instability requires pebble concentration 
 \cite{li2021thresholds,lim2024streaming}, and the inner disk is a favorable region for this, due to inward radial drift of outer disk particles \cite{YoudinShu2002, birnstiel2010gas, drkazkowska2016close}, particle trapping within vortices that form near $r_c$ during infall \cite{Bae2015}, and/or particle buildup \fix{in the outward gas velocity region near $r_c$ \cite{batygin2020formation}} or within a pressure maximum near the silicate evaporation radius \cite{ChatterjeeTan2014}.

Accretion within infall-supplied disks has been studied in the context of gas planet satellite origin, including models that predict a preference for satellite systems with $10^{-4}$ times the planet's mass \cite{CanupWard2006,Cilibrasi2020} (Figure \ref{fig:MassRatio}). A key difference between satellite systems and compact exoplanetary systems is that for the latter, the gas disk persists after the infall ends, which would tend to cause greater migration-driven planet loss.  
The approximate $10^{-4}$ maximum mass ratio seen in compact systems occurs across systems with stellar masses that vary by an order-of-magnitude, suggesting that it does not reflect a maximum absolute planet mass, $M_p$, but instead a maximum $(M_p/M_*)$ value, as would be expected for systems regulated by gas-driven migration whose rate is proportional to $(M_p/M_*)$. 
\subsection*{Planet accretion vs. migration}\label{sec:AnalyticalResults}

We first estimate conditions that allow planets formed during late infall to survive until the gas disk dissipates, and the expected properties of surviving systems.  We consider an infall flux that is radially uniform for $r \le r_c$, has a gas-to-solids ratio $f$, and decreases exponentially with time, $F_{\rm in}=F_0\exp(-t/\tau_{\rm in})$ (where $F_0$ is the flux at the beginning of the final infall stage, and $\tau_{\rm in}$ is related to an observationally determined half-life as $\tau_{\rm in} \sim 1.4t_{1/2}$).  We approximate the gas disk surface density as $\sigma_g(r,t)=\sigma_g(r,0)\exp(-t/\tau_{\rm g})$, where $\tau_g$ is \textcolor{mod}{a gas disk dispersal timescale} and $\sigma_g(r,0)$ reflects an initial quasi-steady state \cite{CanupWard2002} between infall and spreading due to a viscosity $\nu = \alpha cH$, where $\alpha < 1$ is a viscosity parameter, and $c$ and $H$ are the gas sound speed and vertical scale height, respectively.  We define $\beta\equiv(\tau_g/\tau_{\rm in}) > 1$.

Planetesimals form interior to $r_c$ and subsequently grow into larger bodies via mutual collisions and sweep-up of small solids.  
The growth timescale of a planet, $\tau_{\rm acc}$, is regulated by the solid infall rate, with  $\tau_{\rm acc} \sim (f/\varepsilon)M_{\rm p}/(2\pi r\Delta rF_{\rm in})$ \textcolor{mod}{(see Supplemental Discussion \ref{sec:Mtot_final})}, where $\Delta r$ is the planet's feeding zone and $\varepsilon$ is the fraction of infalling solids incorporated into large bodies (either by directly participating in planetesimal formation or by being accreted by large bodies). 

Growing planets undergo inward Type-I migration on a timescale \cite{TanakaWard2004}
\begin{equation}
    \tau_{\rm I}\sim\frac{1}{C_a\Omega(r)}\left(\frac{M_*}{M_{\rm p}}\right) \left(\frac{M_*}{r^2\sigma_g(r,t)}\right)\left(\frac{H}{r}\right)^2
    \label{eq:tauI}
\end{equation}
where $C_a$ is a torque constant, $\Omega(r)=\sqrt{GM_*/r^3}$ is orbital frequency, and $(H/r)\propto (r/R_*)^{\gamma_c}$ with $\gamma_c \ll 1$. 

A planet accretes largely in-place until it approaches a critical mass, $M_{\rm crit}$, for which $\tau_{\rm acc} \sim \tau_{\rm I}$ \cite{CanupWard2006}. The planet then migrates inward before becoming substantially more massive.  Setting $\tau_{\rm acc} = \tau_{\rm I}$ and solving for $M_{\rm p}=M_{\rm crit}$ gives

 \begin{equation}
 \begin{aligned}
     \frac{M_{\rm crit}(t)}{M_*}    &\approx 4.1 \left(\frac{M_*}{\pi r_c^2F_0} \frac{f}{\varepsilon}\Omega(r_c)\right)^{1/9}\left(\frac{\pi}{C_a}\right)^{5/9}  \left(\frac{H}{r}\right)^{26/9}\left(\frac{r}{r_c}\right)^{17/18} \left(\frac{\alpha\varepsilon}{f}\right)^{2/3}\\
  & \ \ \ \ e^{ -t/\tau_{\rm in}\left( \frac{5}{9}-\frac{2}{3\beta}\right)}\\   
  &\approx 2.8\times10^{-5}\chi\left(\frac{1}{C_a}\right)^{5/9}\left(\frac{H/r}{0.05}\right)^{26/9}
  \left(\frac{r}{r_c}\right)^{17/18} \left(\frac{\alpha\varepsilon/f}{5\times10^{-5}}\right)^{2/3} \\
  & \ \ \ \ e^{-t/\tau_{\rm in}\left( \frac{5}{9}-\frac{2}{3\beta}\right)}
 \end{aligned}
 \label{eq:Mcrit}
\end{equation}

where $\chi\equiv\left(\frac{M_*/(\pi r_c^2F_0)}{16\ \rm{Myr}}\frac{f/\varepsilon}{100}\frac{100\ \rm{days}}{2\pi/\Omega(r_c)}\right)^{1/9}$ is a term of order unity \textcolor{mod}{that depends only weakly on disk and inflow parameters}. 

As a mass $M_{\rm crit}$ planet migrates inward, a new mass $M_{\rm crit}$ planet can form in its place on a comparable timescale due to ongoing solid infall. This regulates the total mass in planets, $M_{\rm tot}$, to (Supplemental discussion \ref{sec:Mtot_final}) 
\begin{equation}
 \begin{aligned}
    \frac{M_{\rm tot}(t)}{M_*} &= \int^{r_c}_{r_{\rm in}}\frac{M_{\rm crit}/{M_{*}}}{\Delta r} \rm{d}r \\
    &\approx  2.2\times10^{-4}\frac{1}{\chi}\left(\frac{1}{C_a}\right)^{4/9}\left(\frac{\alpha\varepsilon/f}{5\times10^{-5}}\right)^{1/3}\left(\frac{H/r}{0.05}\right)^{10/9}e^{-t/\tau_{\rm in}\left(\frac{4}{9}-\frac{1}{3\beta}\right)}
 \label{eq:Mtot}
 \end{aligned}
\end{equation}
where $r_{\rm in}$ is an inner loss boundary (we set $r_{\rm in}/r_c\sim0.1$), and we neglect the weak dependence of $(H/r)$ on $r$ given that $\gamma_c  \ll 1$. 

The predicted system mass ratio in eqn. \ref{eq:Mtot} depends very weakly on the infall rate through $\chi$.  Of importance is the  $(\alpha\varepsilon/f)^{1/3}$ term.  The fraction of the total infalling mass incorporated into solid planets is $(\varepsilon/f)$, and if this fraction is increased for a fixed $\alpha$, accretion becomes faster and planets grow more massive before migrating inward. For a fixed $(\varepsilon/f$), a higher $\alpha$ leads to faster disk-spreading and a lower gas surface density, which slows migration and allows for more massive planets. In addition (and differently from satellite origin models, \citep{CanupWard2006}), there are time-dependent terms in eqns. \ref{eq:Mcrit} and \ref{eq:Mtot} due to the mismatch between the infall timescale, $\tau_{\rm in}$, and the gas disk \textcolor{mod}{dispersal timescale}, $\tau_g$.  

The balance between accretion and migration described by eqns. \ref{eq:Mcrit} and \ref{eq:Mtot} continues until the migration time for a mass $M_{\rm crit}$ planet becomes much longer than the gas disk \textcolor{mod}{dispersal timescale}, i.e., until $\tau_{\rm I}(M_{\rm crit}) \gg \tau_{\rm g}$. As this transition occurs, migration subsides and mass $M_{\rm crit}$ planets can survive. We determine the time at which a mass $M_{\rm crit}$ planet at $r_c$ has $\tau_I = 10\tau_g$, and evaluate eqn. \ref{eq:Mtot} at this time to estimate the final planetary system mass ratio that survives against Type-I migration (Supplemental discussion \ref{sec:Mtot_final}):  

\begin{equation}
\begin{aligned}
    \frac{M_{\rm tot,final}}{M_*} 
    & \approx \left(\frac{M_{\rm tot,0}}{M_*}\right)^{1+\Lambda}\left(\frac{1}{2}\frac{M_*}{M_{\rm in}}\frac{f/\varepsilon}{10\beta}\right)^\Lambda
    \label{eq:Mtot_Final}
\end{aligned}
\end{equation}
where $(M_{\rm tot,0}/M_*)$ is a maximum value from eqn. \ref{eq:Mtot} for $t \ll \tau_{\rm in}$, $\Lambda\equiv(4\beta-3)/(5\beta+3)$ with $0.125\le \Lambda \le 0.8$  for $1\le \beta \lt \infty$, and $M_{\rm in} \equiv \pi r_c^2F_0\tau_{\rm in}$ is the total mass delivered during the final infall stage.  As $\beta$ increases, the surviving system mass ratio decreases.  

\subsection*{Numerical simulations}\label{sec:Results}

We simulate this process using an N-body code \cite{Duncan1998} that includes a time-dependent infall of gas and solids, and gas disk torques (see \nameref{sec:Nbodysims} section in Methods). 
\fix{There is strong observational evidence for inner magnetospheric cavities in the disks of Class II (post-infall) stars \cite{romanova2006magnetospheric,zhu2023global}, and such a cavity can halt migration of a planet near its edge (see Supplemental discussion \ref{sec:InnerDisk cavity}). Whether disks of Class I stars undergoing infall have inner cavities is less clear, as they may instead undergo boundary layer accretion without a  cavity \cite{Gaches2024}.   Given this, we consider two limiting cases:  a disk with an inner migration-stopping cavity, and a no-cavity disk with full inward migration.}  

\begin{figure}
     \centering
     \includegraphics[width=\textwidth]{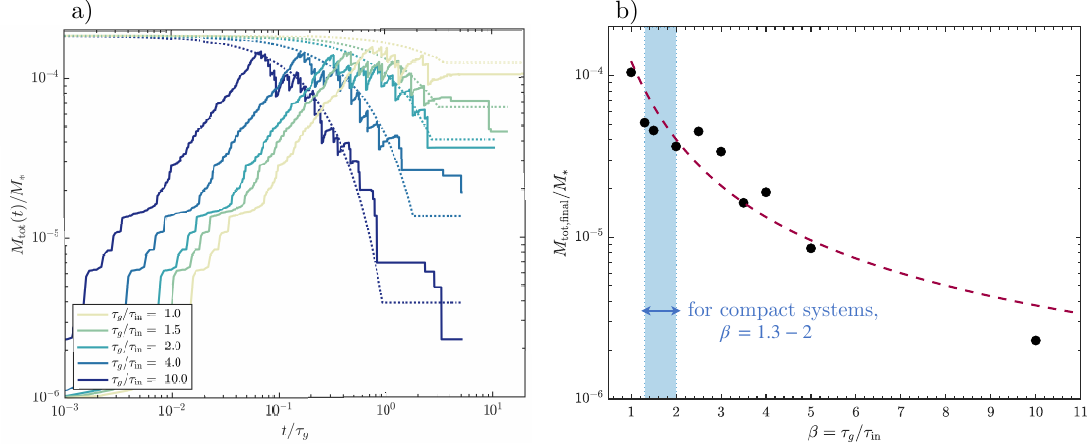}
     \caption{\textbf{Total planetary system mass during infall and subsequent gas disk dissipation, in units of the stellar mass}. \textbf{a,} System mass ratio vs. time normalized by the gas disk dispersal timescale, $\tau_g$. Simulations here assume $(\alpha \varepsilon/f)=5\times10^{-5}$ (where $\alpha$ is the disk viscosity parameter, $\varepsilon$ is the fraction of infalling solids incorporated into planets, and $f$ is the infall gas-to-solids ratio), a common infall prescription, no inner cavity (i.e., inward migration of all planets), and varied  $\beta \equiv \tau_g/\tau_{\rm in}$  values indicated by color (where $\tau_{\rm in}$ is the infall decay timescale). Solid lines are simulation results; dotted curves show predictions from eqns. \ref{eq:Mtot} and \ref{eq:Mtot_Final}.  \textbf{b,} Final system mass ratio vs. $\beta$ from 10 simulations compared to analytical prediction (eqn. \ref{eq:Mtot_Final}, red dashed line, assuming $f /\varepsilon = 10^2$). Blue shaded region shows the range appropriate for compact systems with small $r_c$, although planetary systems survive even for larger $\beta$ values.}
     \label{fig:MassTime_SameInfall}
 \end{figure}

Figure \ref{fig:MassTime_SameInfall} shows results of simulations \textcolor{mod}{with full migration} that consider different $\beta$ for otherwise equal conditions. 
The system mass initially increases due to the solid infall until planets of mass  about $M_{\rm crit}$ form and migrate inward.  As planets are lost by migration, new planets accrete in their place as the infall and gas disk properties evolve.  The total planetary system mass (Figure \ref{fig:MassTime_SameInfall}a, solid lines) oscillates around a value that decreases with time, as predicted by eqn. \ref{eq:Mtot} (Figure \ref{fig:MassTime_SameInfall}a, dotted lines), until the transition time at which $\tau_I(M_{\rm crit})\gg\tau_g$, after which the total mass stabilizes. As the relative \textcolor{mod}{disk dispersal timescale} compared to the infall timescale increases (varied colors in Figure \ref{fig:MassTime_SameInfall}a), the final system mass ratio decreases, consistent with eqn. \ref{eq:Mtot_Final} (Figure \ref{fig:MassTime_SameInfall}b).

To constrain the appropriate values for $\beta$ for compact systems, we model the radial evolution of a gas disk subject to a time-dependent infall, viscous spreading, and photoevaporative loss. 
Disks with smaller $r_{\rm c}$ have infalling material deposited at smaller radii where viscous timescales are shorter and photoevaporative losses can be greater, leading to shorter disk lifetimes \cite{HuesoGuillot2005,schib2021influence}.  We find that for $r_c < 1$ au, $\alpha >$ few $\times 10^{-4}$, and $t_{1/2} > 2 \times 10^5$ yr (i.e., $\tau_{\rm in} \ge 3 \times 10^5$ y), the gas disk's evolution may be approximated as having $1.3 \le \beta \le 2$ (see \nameref{sec:GasEvolutionSup} section in Methods). For this range of $\beta$, eqn. \ref{eq:Mtot_Final} implies that surviving systems formed during infall would have  about $20-40\%$ of $M_{\rm tot,0}$ (assuming $f/\varepsilon=10^2$ and $M_{\rm in}/M_*=0.03$ as in Figure \ref{fig:MassTime_SameInfall}-b). \fix{The higher $\beta$ values associated with compact disks compared with circumplanetary disks (thought to have $\beta \sim 1$ \cite{CanupWard2002}) would lead to a somewhat lower mass ratio for compact systems compared to satellite systems, which appears the case (Figure \ref{fig:MassRatio}).}  \textcolor{mod_ver3}{We note, however, that there is not a sharp cutoff as $\beta$ increases, and compact systems survive even with $\beta > 2$ (Figure \ref{fig:MassTime_SameInfall}-b).} 

\begin{figure}
      \centering
     \includegraphics[width=0.65\textwidth]{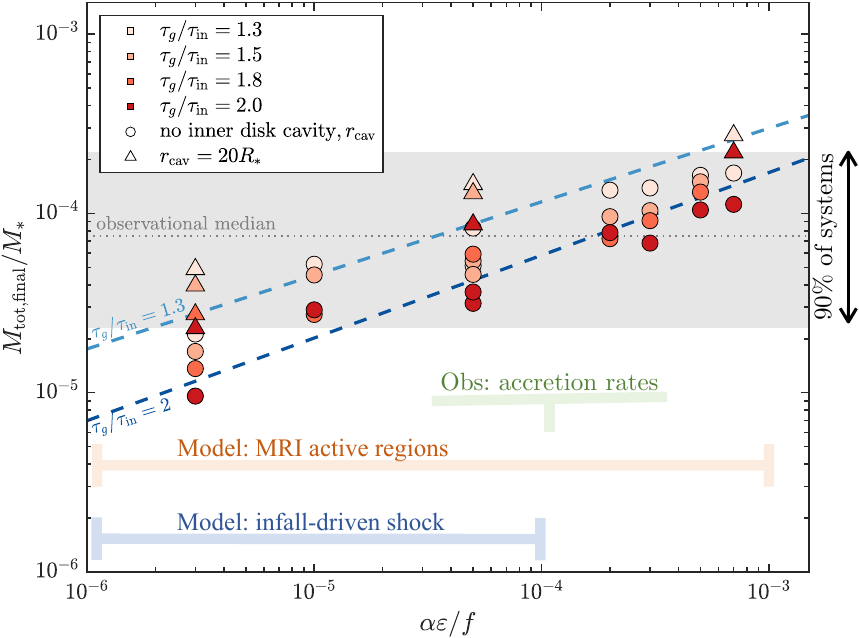}
     \caption{\textbf{Results of planet accretion simulations with varied disk and infall properties}. Final planetary system mass scaled to the stellar mass ($1M_\odot$) as a function of $(\alpha\varepsilon/f)$  ($\alpha$ is the viscosity parameter, $\varepsilon$ is the fraction of infalling solids incorporated into planets, and $f$ is the infall gas-to-solids ratio). The infall rate decays with timescale $\tau_{\rm in}= 5 \times 10^5$ yr, while the gas disk disperses over a longer timescale, $\tau_g=$ 1.3 to 2$\tau_{\rm in}$ (colors, legend). The simulations assume either an inner disk cavity with radius $r_{\rm cav}=20 R_*\sim0.13\,\rm{au}$ (triangles) or no cavity (circles). Grey region shows range for 90\% of observed compact systems shown in Figure \ref{fig:MassRatio}. Dashed lines show eqn. \ref{eq:Mtot_Final} predictions for the no-cavity case with $\beta=1.3$ (light blue) and $\beta=2$ (dark blue), with $H/r=0.05$ and $M_{\rm in }/M_*=0.03$. Horizontal bars show plausible viscosity ranges based on observed stellar accretion rates \cite{Hartmann1998}, models of magnetorotational instability (MRI) \cite{jankovic2019close} and infall-driven shocks \cite{Lesur2015}, assuming $f/\varepsilon = 10^2$.} 
     \label{fig:MassAlpha}
 \end{figure}

Figure \ref{fig:MassAlpha} shows results of simulations with varied $(\alpha/\varepsilon f)$ and $1.3 \le \beta \le 2$.  
\fix{For both disks with and without inner cavities,} the final system mass ratio is between $10^{-5}$ and $2\times 10^{-4}$ \fix{across three orders-of-magnitude variation in $(\alpha\varepsilon/f)$}. \textcolor{mod}{\fix{For otherwise similar conditions, simulations with an inner cavity (triangles) yield systems with somewhat larger mass ratios than no-cavity cases (circles) and the no-cavity analytical estimate (eqn. \ref{eq:Mtot_Final}, dashed lines)}. When a planet nears the cavity edge, its positive co-rotation torque may become comparable to its negative Type-I torque so that its gas disk interactions no longer cause it to migrate \cite{masset2006disk}.  The inner planet may then hold back the migration of exterior planets via capture into mutual resonances \fix{\cite{Izidoro2017}}, allowing the \fix{resulting resonant chain of} planets to grow somewhat more massive than in the no-cavity case.  However, \fix{the barrier to planet infall is not absolute}.  As exterior planets continue to grow due to the infall, the inner planet's co-rotation torque becomes insufficient to offset the outer planets' negative Type-I torque, and the chain of planets collapses inward.
\fix{As a result,} the planet accretion-migration cycle, and its dependencies on disk and infall parameters, are broadly similar in the no-cavity and cavity cases, although a cavity allows a given mass ratio system to be achieved with somewhat less restrictive parameters, i.e., lower $(\alpha\varepsilon/f)$ and/or higher $\beta$ (see Supplemental discussion \ref{sec:InnerDisk cavity}).}

The appropriate values for $\varepsilon$ and $\alpha$ are highly uncertain, and could vary across different systems.  \textcolor{mod}{However, per eqn. \ref{eq:Mtot_Final}, the final system mass is not strongly dependent on these parameters: for $1 < \beta \le 2$, $(M_{\rm tot,final}/M_*)\propto \alpha^{3/8}$ to $\alpha^{1/2}$, and $(M_{\rm tot,final}/M_*)\propto \varepsilon^{1/13}$ to $\varepsilon^{1/4}$.} \fix{The infall metallicity ($1/f$)} would nominally be comparable to the stellar metallicity. All other factors being equal, a higher infall metallicity (smaller $f$) leads to  \textcolor{mod}{more massive planets and a larger system mass ratio, although here too the dependence is weak, with $(M_{\rm tot,final}/M_*) \propto f^{5/9\Lambda - 4/9}$, so that, e.g.,  
a factor of $2$ metallicity enhancement yields only a factor of 1.2 to 1.3 increase in mass ratio for $1.3\le \beta \le 2$. This \fix{appears consistent with} the nearly equal occurrence of warm super-Earths around stars of wide-ranging metallicities \cite{petigura2018california} that has been unexplained. }  

\begin{figure}
      \centering
     \includegraphics[width=0.9\textwidth]{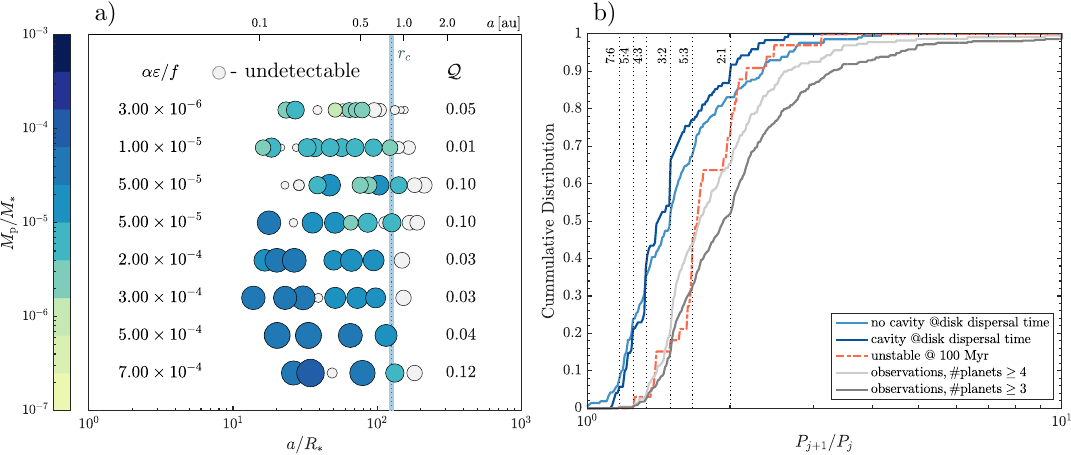}
     \caption{\textbf{Predicted system architectures and period distributions}. \textbf{a,} Example simulation results for a $1 M_\odot$ star, full inward migration, and $\beta=1.3$ (see Supplemental discussion \ref{sec:InnerDisk cavity} and \ref{sec:TrapSystem} for all results including inner cavity cases).   Systems are arranged by increasing $(\alpha\varepsilon/f)$ ($\alpha$ is the viscosity parameter, $\varepsilon$ is the fraction of infalling solids incorporated into planets, and $f$ is the infall gas-to-solids ratio). Symbol color and size scale with planet mass ratio, ($M_p/M_*$); grey indicates planets estimated to be undetectable (Supplemental discussion \ref{sec:DetectabilitySup}); and planets orbiting within 3 mutual Hill radii have been merged.  The outer distance at which planetesimals form (assumed here to be equal to $r_c$, blue vertical line) establishes the radial scale of the planet system, although growing planets can be scattered to somewhat larger distances too. The mass partitioning parameter, $\mathcal{Q}$, for each system is similar to those of observed systems that have $\mathcal{Q} < 0.2$. \textbf{b,} Period ratio distribution of adjacent planets in simulated systems versus observed systems from Figure \ref{fig:MassRatio}, including locations of mean motion resonances (vertical dashed lines). The dark [light] blue curve shows the period distribution soon after gas disk dispersal for simulations with an inner cavity [without a cavity]. The red dashed curve shows this distribution for systems that underwent a dynamical instability during longer, $100$ Myr orbital integrations. As in prior works \cite{Izidoro2017,Izidoro2021}, we find that later instability leads to improved agreement between model results (red dashed curve) and observations (dark and light grey curves).}
     \label{fig:Planetary Systems} 
    
 \end{figure}

Figure \ref{fig:Planetary Systems}a shows resulting planet distributions at 5 to 10 Myr from a subset of simulations \textcolor{mod}{with full migration}.  Observed compact systems have similar planet sizes within each system, a ``peas-in-a-pod'' architecture. \fix{In our model, this similarity emerges naturally through} the regulation of planet mass by Type-I migration (eqn. \ref{eq:Mcrit}). Size diversity can be quantified by $\sigma_\mathcal{R}^2= {\rm Var}\left[{\rm log}_{10}\left(R_p/R_\oplus\right)\right]$ \cite{weiss2022architectures}, where $(R_p/R_\oplus)$ is planetary radius scaled to Earth's radius\textcolor{mod_ver2}{, or with a mass partitioning coefficient, $\mathcal{Q}=N/(N-1)\sum_{i=1}^{N}(m_p/M_{\rm tot}-1/N)^2$ \cite{gilbert2020information}}. Known multi-planet compact systems have a median $\sigma_\mathcal{R}=0.14$ \cite{weiss2022architectures},
with most systems having $\sigma_\mathcal{R}=0.08$ to $0.19$\textcolor{mod_ver2}{, and $\mathcal{Q}<0.2$}.  Our simulated systems (both with and without an inner cavity; \fix{see Supplemental discussion \ref{sec:InnerDisk cavity} and \ref{sec:TrapSystem}}) compare favorably to these\fix{, having $\mathcal{Q} < 0.2$ (Figure \ref{fig:Planetary Systems}a) and a median value of $\sigma_\mathcal{R} \approx 0.16$ (assuming 
a mass-radius scaling law \cite{WeissMarcy2014}).
As is also the case for prior (post-infall) migration models \cite{Izidoro2017,Izidoro2021}, our systems soon after gas disk dispersal have, on average, a greater number of planets (typically 5 to 7 with $M_{\rm p} >10^{-6}M_*$) than observed systems (Figure \ref{fig:Planetary Systems}a).}

\textcolor{mod_ver3}{Another key comparison is the planet period distribution.  Planets in observed compact systems are not predominantly in mean motion resonances (Figure \ref{fig:Planetary Systems}b, dark and light grey lines) \cite{fabrycky2014architecture}.  In contrast, planet formation models involving migration (including ours and prior, post-infall models \cite{Izidoro2017,Izidoro2021}) predict more resonant pairs and generally more compact systems near the time of gas disk dispersal.  Our cases without an inner cavity (Figure \ref{fig:Planetary Systems}b, light blue curve) show fewer resonant pairs than cases with a cavity (Figure \ref{fig:Planetary Systems}b, dark blue curve), because the cavity favors the formation of resonant chains, but both distributions are initially distinct from observations (Figure \ref{fig:Planetary Systems}b, grey curves).} 

A strong case has been made that on longer timescales, dynamical instabilities cause the period ratio distribution to evolve and become more consistent with the observed distribution \cite{Izidoro2017,Izidoro2021}. To illustrate such an evolution, we performed 100 Myr integrations of our systems assuming the gas disk had completely dispersed.  As in prior works \cite{Izidoro2017,Izidoro2021}, we find that systems that undergo instability have a period ratio distribution that more closely matches that of observed systems (Figure \ref{fig:Planetary Systems}b, red curve vs. grey curves).  The system mass ratios are not substantially affected by this later evolution.  Often instability involves planet merger (which leaves the system mass ratio unchanged), and in cases where planets are ejected, their mass is typically a small fraction of the total system mass.  Of the systems that underwent instability in our 100 Myr integrations, 90\% had their mass ratio change by $< 5$\%.  As argued previously, continued interactions on longer timescales and/or other effects \cite{matsumoto2020breaking} may drive continued evolution of the period distribution and further decrease the number of observable planets \cite{Izidoro2017,Izidoro2021}.

\section*{Discussion}
It was proposed \cite{batygin2023formation} that compact system planets accrete within a narrow ring of planetesimals at the silicate condensation distance \fix{ and then migrate inward until they are stopped by an inner cavity} \cite{morbidelli2022contemporary}. \textcolor{mod}{The Ref \cite{batygin2023formation} model is similar to ours in that inward Type-I migration produces a uniformity in planet mass within each system. \fix{A primary difference is that Ref }
\cite{batygin2023formation} adopts a smaller $r_c$, so that the silicate condensation line and planetesimal formation occur exterior to $r_c$ where the disk is not directly supplied by infall.  The final planet system mass in \cite{batygin2023formation} is then set by the assumed mass of the outer planetesimal ring. Why this initial ring mass would have the value needed to produce a preferred system mass ratio across varied stellar masses is not addressed. In contrast, we consider that planetesimals and planets form interior to $r_c$ in the region of ongoing infall.  It is this condition that leads to a common system mass ratio (eqn. \ref{eq:Mtot_Final}).}

\fix{Predictions of our model should be increasingly testable by observations.  Accretion during infall implies a common compact system mass ratio independent of stellar mass, and that systems with somewhat lower estimated mass ratios might host yet undetected planets.  An unusually weak dependence of planet and planet system mass on stellar metallicity is predicted, with $(M_{\rm tot}/M_*)$ proportional to metallicity to the 0.2 to 0.3 power, in contrast to other models that generally imply a stronger dependence.  While we here considered planet accretion interior to $r_c$, some small solids would be carried beyond $r_c$ by expanding gas and would be available for accretion in that region.  Results here imply a difference between compact system planets formed interior to $r_c$ -- that display a characteristic planet mass and system mass ratio -- and those that might form beyond $r_c$, whose masses would not be regulated in the same manner. Such a transition in planet mass ratio might be detectable.}

While the masses and orbital properties of compact systems formed during infall prove to be insensitive to disk conditions (e.g., the specific value of $r_c$, infall rate, and viscosity), resulting planet compositions will be more sensitive to such parameters. Our model argues for rocky planet formation interior to the ice line \cite{LuquePalle2022,batygin2023formation}. As infall slows, the disk cools \cite{CanupWard2002} and condensation fronts shift inward.  While early-formed planets would tend to be refractory rich (and more likely to be lost), some conditions (e.g., small $\alpha$, larger $r_c$) might allow for ice stability during late infall.  Coupled composition and dynamical models will be needed to assess this.

Overall, whether planets that form during infall survive, and what the characteristics of surviving planets would be, will depend on $r_c$ (the radial extent of disk infall) and $\beta$ (the ratio of the gas disk lifetime to the infall timescale).  Here we focus on small $r_c$ and small $\beta$ cases, consistent with compact systems.  However, $r_c$ depends sensitively on the angular momentum of the parent cloud core and interactions of infalling material with the disk and stellar magnetosphere, so that even stars with similar masses may have substantially different $r_c$ values. For larger $r_c$, long accretion timescales in the outer infall region may promote the formation of giant planets after infall has ended.  Larger $r_c$ systems will also have larger $\beta$, due to their longer gas disk lifetimes \cite{schib2021influence}, implying that surviving inner planets that accreted during infall would have a lower total mass compared to the star; e.g., per Figure \ref{fig:MassTime_SameInfall}, a mass ratio of inner planets similar to our terrestrial planets might result for $\beta\sim 8$. Most broadly, our results suggest that the long-standing assumption that planet accretion commences only after infall has ended may not be valid for all systems, and consideration of this early accretionary phase is warranted.


\section*{Methods}\label{sec: Methods}

\subsection*{Relative timescales of gas disk dispersal vs. infall}\label{sec:GasEvolutionSup}

We show that the properties of the final planetary system formed during infall depend on the ratio of the gas disk dispersal timescale $\tau_g$, to the timescale over which disk infall ends, $\tau_{\rm in}$ (eqn. \ref{eq:Mtot_Final} and Figure \ref{fig:MassTime_SameInfall}). To estimate the range of $\beta \equiv \tau_g/\tau_{\rm in}$ appropriate for accretion disks with small $r_c < 1$ au, we model the evolution of a gas disk subject to a time-dependent infall, viscous spreading, and photoevaporative loss \cite{Owen2012} using a semi-analytical code that tracks the gas surface density, $\sigma_{g}(r,t)$  as the last $10\%$ of the stellar mass infalls to the disk and continues through the gas disk dispersal phase. 

We consider a uniform infall flux per disk area, $F_{\rm in}$, that decays exponentially with timescale $\tau_{\rm in}$, with $10^5 < \tau_{\rm in} < 10^6$ yr. We consider an $\alpha$-viscosity model with $\nu = \alpha c H = \alpha c^2/\Omega$ (where $c$ and $H$ are the gas sound speed and vertical scale height, and $\Omega$ is the orbital frequency) and few $\times 10^{-4} < \alpha < 0.1$. We adopt typical X-ray luminosity values ($L_X=10^{29} \rm {\ erg\ s^{-1}}$ for a mass $0.1M_\odot$ star and $L_X=10^{30} \rm {\ erg\  s^{-1}}$ for a $1M_\odot$ mass star \cite{preibisch2005evolution}). 

To set the starting gas surface density and radial temperature profile, we assume that the disk is initially in a quasi-steady-state between infall and viscous spreading, where the latter depends on temperature in the $\alpha$-model \cite{CanupWard2002}. At each disk radius $r$, we solve for the mid-plane disk temperature, $T$, that balances: 1) radiative cooling from the disk’s top and bottom surfaces; 2) heating due to the star’s luminosity with $\dot{E}_*=9/4\sigma_{\rm SB}T_*^4(R_*/r)^2(c/r\Omega)$ (where $\sigma_{\rm SB}$ is the Stephan-Boltzmann constant and $T_*$ is the stellar temperature); 3) heating due to viscous dissipation, $\dot{E}_\nu = (9/4)\sigma_{g}\nu\Omega^2$; and 4) dissipation due to the energy difference between infalling gas and that of a local keplerian orbit, $\dot{E}_{\rm in} \sim F_{\rm in}G M_*/r$ \cite{nakamoto1994formation}:

\begin{equation}
    T^4 = \left(1+\frac{3\kappa_R\sigma_{g}}{8}+\frac{1}{2\kappa_P\sigma_{g}}\right){{\dot{E}_\nu}\over{2\sigma_{\rm SB}}}+
    \left(1 + \frac{1}{2\kappa_P\sigma_{ g}}\right){{\dot{E}_*+\dot{E}_{\rm in}}\over{2\sigma_{\rm SB}}}+T^4_{cd}
    \label{eq:Temp}
\end{equation}
where $\kappa_R$ [$\kappa_P$] is the disk's frequency-averaged Rosseland [Plank] opacity \cite{nakamoto1994formation} and $T_{\rm cd}=15$ K is the molecular cloud temperature. This expression assumes that heating due to stellar luminosity and infalling material is deposited at the disk surface, while viscous dissipation occurs primarily in the mid-plane \cite{nakamoto1994formation}. Given $T(r,0)$, we calculate the initial viscosity and surface density ($\nu(r,0)$ and $\sigma_{\rm g}(r,0)$) from eqn. (18) of ref. \cite{CanupWard2002}.

\begin{figure}
    \centering
    \includegraphics[width=0.7\textwidth]{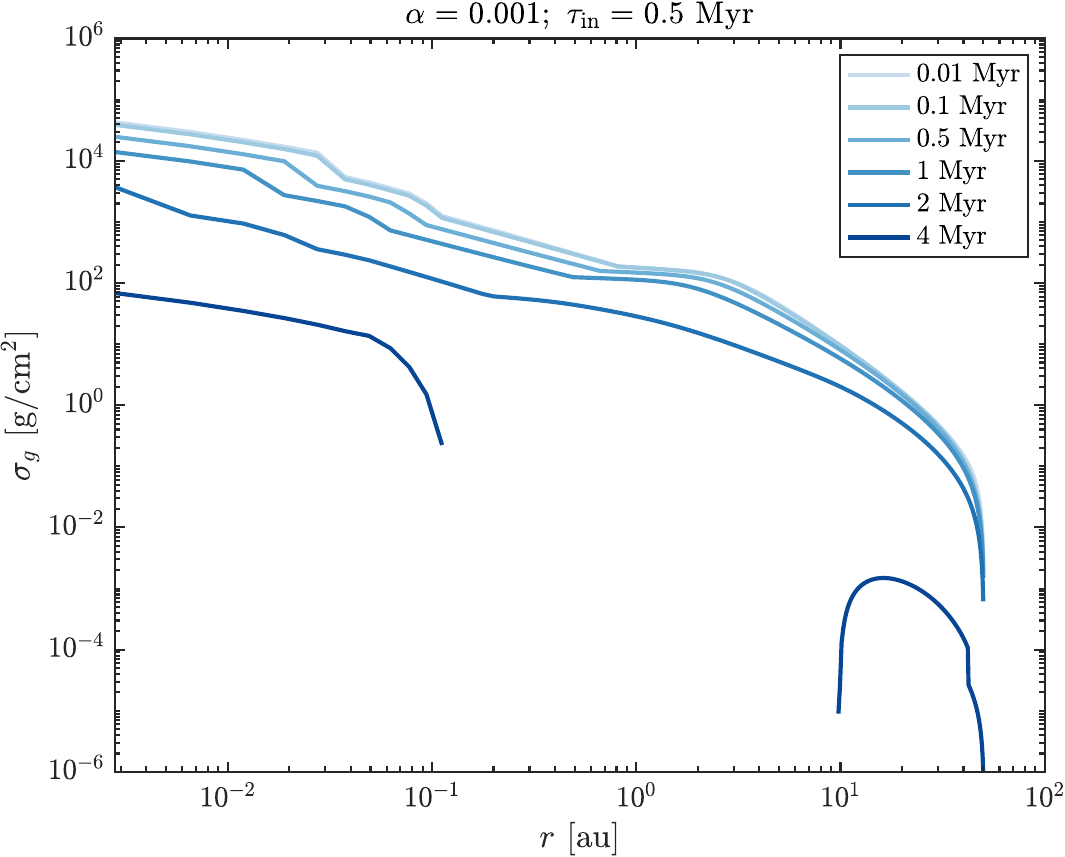}
    \caption{\textbf{Gas disk surface density evolution}. Example evolution of the gas disk surface density around a $0.1M_\odot$ star at different times with $\alpha=10^{-3}$, $r_c = 0.06$ au, and $\tau_{\rm in}=0.5$ Myr.  By 4 Myr, a large gap has formed due to photoevaporative losses.}
    \label{fig:DensityOverTime}
\end{figure}

To calculate $\sigma_g(r,t)$, we use standard methods for a viscous disk \cite{BathPringle1981} subject to photoevaporative loss \cite{Owen2012,preibisch2005evolution}, assuming vertical thermal balance in the disk at each radius per eqn. \ref{eq:Temp}. Figure \ref{fig:DensityOverTime} shows an example evolution.

\begin{figure}
    \centering
    \includegraphics[width=1\textwidth]{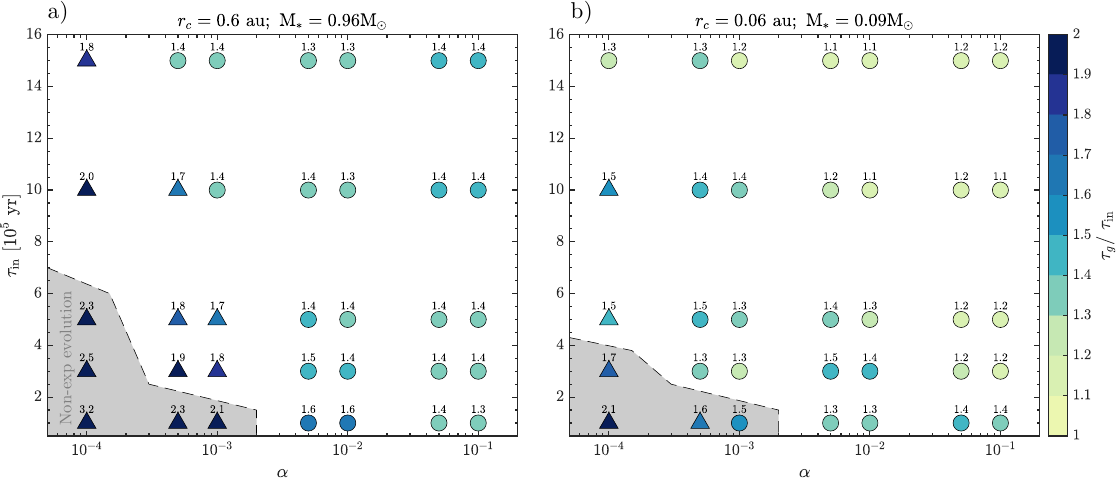}
    \caption{\textbf{Gas disk lifetime compared to infall timescale for compact systems}. Predicted ratio ($\beta$) between the gas disk lifetime, $\tau_g$ and the infall timescale, $\tau_{\rm in}$, as a function of the disk viscosity parameter ($\alpha$) and $\tau_{\rm in}$ for compact systems for a \textbf{a}) solar and \textbf{b}) sub-solar  mass star.  Resulting values of $\beta \equiv\tau_g/\tau_{\rm in}$ are shown by color and numerical labels. Circles [triangles] show cases with initial gas disk masses $<0.1M_*$ [$>0.1M_*$]. Simulations depicted by triangles assumed an initial viscosity of $\alpha_0=0.1$, replicating the expected high-viscosity for gravitationally unstable, high-mass disks. Simulations with very low viscosities and fast infall, depicted by the gray-shaded regions in the lower-left corners of the plots, exhibit a disk mass time-evolution that is not well approximated by a single exponential. For cases outside this region, the evolution of the inner gas disk mass can be reasonably well approximated as proportional to $e^{-(t/\tau_{\rm g})}$ (see Figure \ref{fig:MdiskVsTime}).}
    \label{fig:TaugTauIn}
\end{figure}

For low viscosity cases with $\alpha =10^{-4}$, the initial disk mass estimated in this manner is 
$>0.1M_*$.  There is then an inconsistency, because such a massive disk would have a large, gravitationally-driven viscosity with $\alpha \sim O(10^{-1})$ until its mass had decreased substantially \cite{kratter2016gravitational}.  For such cases (indicated by the triangle markers in Figure \ref{fig:TaugTauIn}), we set $\alpha = 0.1$ when calculating $T(r,0)$ and $\sigma_g(r,0)$, and
then proceed with $\alpha = 10^{-4}$ to calculate the post-gravitational instability evolution.  

For each disk evolution with a given $\alpha$ and $\tau_{\rm in}$, we calculate an effective $\tau_g$ as the time needed for the gas mass interior to $r_c$ to decrease by $1/\exp$. \textcolor{mod}{Note that because $\tau_g$ is an \textit{e}-folding time, not the disk lifetime, significant gas is still present at $t = \tau_g$.} In general, $\tau_g$ decreases [increases] for larger [smaller] viscosity $\alpha$-values. 

\begin{figure}
    \centering
    \includegraphics[width=1\textwidth]{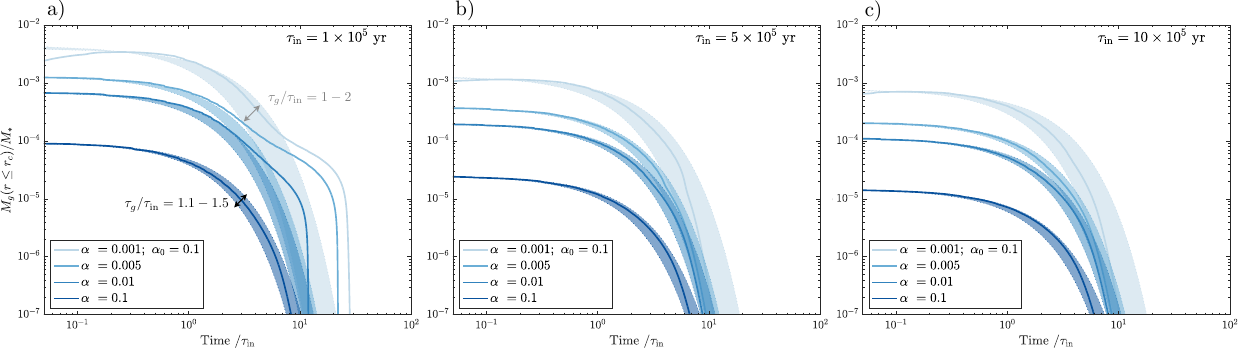}
    \caption{\textbf{Gas disk mass evolution over time for different infall timescales}. The gas disk mass interior to $r_c$ scaled to the stellar mass as a function of time normalized by the infall timescale, for \textbf{a)} $\tau_{\rm in}=3\times10^5$ yr; \textbf{b)} $\tau_{\rm in}=5\times10^5$ yr; and \textbf{c)} $\tau_{\rm in}=10\times10^5$ yr. The colored curved lines correspond to different viscosity values, per the legend, while the shaded area surrounding each curve shows the inner disk mass evolution predicted by an exponential decay with $\beta$ between 1 and $2$ for the $\alpha=0.001$ cases, and with $\beta$ between 1.1 and 1.5 for the other viscosity values. For disks produced by infall with small $r_c < 1$ au, the effective $\beta$ values are generally quite low. }
    \label{fig:MdiskVsTime}
\end{figure}

We use this model to estimate $\beta \equiv \tau_g/\tau_{\rm in}$ as a function of $\alpha$ and $\tau_{\rm in}$.  Across a broad parameter range, specifically for $\tau_{\rm in} \ge 3 \times 10^5$ yr and $\alpha >$ few $\times 10^{-4}$, we find that $\beta$ is between 1.3 and 2 for compact systems with $r_c < 1$ au (Figure \ref{fig:TaugTauIn}a-b, non-shaded regions). Figure \ref{fig:MdiskVsTime} shows the mass of gas interior to $r_c$ vs. time for example cases in this parameter range.  The gas decrease is reasonably well approximated by an exponential decay with \textcolor{mod}{timescale $\tau_g\sim1.3\,\tau_{\rm in}$ to $2\,\tau_{\rm in}$}. \textcolor{mod}{While $\beta$ is small for these compact disks, resulting gas disk lifetimes (defined as when the optical depth in the near-infrared region, $T=300$ K, is smaller than unity) are $>3$ Myr (for $\tau_{\rm in} \ge 3 \times 10^5$ yr and $\alpha >$ few $\times 10^{-4}$), in agreement with previously studies \cite[e.g.,][]{schib2021influence}.}

For cases with low viscosities and fast infall rates, the gas disk evolution is not well approximated as a simple exponential decay; this regime is indicated by the gray-shaded region in Figure \ref{fig:TaugTauIn}. For these cases, the late disk mass decreases more slowly than implied by a single exponential, and the resulting disk masses are large, comparable to the mass of the star. 
 More sophisticated models are required to accurately estimate disk evolution in this regime \cite{kratter2016gravitational}.
 
\subsection*{$N$-body planet accretion simulation}\label{sec:Nbodysims}

To simulate planet accretion during and after infall, we use a modified version of SyMBA \cite{Duncan1998}, a symplectic integrator that allows for close encounters and collisions (treated as inelastic mergers). The background gas disk is treated analytically and its effects are added as small perturbations to the planetary orbits 
\cite{CanupWard2006}. We continue the simulations for 5 to 10 times the gas disk dispersal timescale, by which time migration-driven effects have ended.  Collisions among planets may occur on longer timescales, which would be expected to alter somewhat the final architecture of systems shown in Figure \ref{fig:Planetary Systems}. As in other $N$-body models of compact system origin, \cite[e.g.,][]{OgiharaIda2009,Cossou2014,Izidoro2017,batygin2023formation}, we do not include gas accretion by the growing planets, given that the planets in these systems are mostly solid by mass. 

\subsubsection*{$N$-Body Gas Disk Description}  \textcolor{mod}{Our accretion simulations adopt a simplified gas disk treatment based on results of the 1-D disk model described above}.  We assume $\sigma_g(r,t)$ varies as a power-law in orbital radius with a constant power-law exponent ($\gamma_g$), and that the gas dissipates on an exponential timescale, $\tau_g$, so that $\sigma_{g}(r,t)=\sigma_{g}(0,R_*)(R_*/r)^{\gamma_g}{\rm exp}(-t/\tau_g)$. We set $\gamma_g = 0.65$ and $\sigma_g(0,R_*)\propto F_{\rm in}/\nu$ to approximate an initial quasi-steady state with a disk opacity of order unity \cite{CanupWard2002}. The disk vertical aspect ratio is set as $(H/r) =  (H/r)_{R_*}(r/R_*)^{\gamma_c}$ with $\gamma_c = 0.04$ (implying a temperature profile $T(r) \propto r^{2\gamma_c-1}$), and we make the simplifying assumption that $(H/r)$ is constant with time. For simulations of accretion around a solar mass star in the main text, we set $\gamma_c = 0.04$, and $(H/R)_{R_*} = 0.04$; additional simulations for a more compact, Trappist-like system with $M_* = 0.1M_\odot$ set $(H/R)_{R_*} = 0.06$ (see Supplemental discussion \ref{sec:TrapSystem}).  

Interactions with the gas disk cause orbiting bodies to undergo inward Type-I migration (with timescale from eqn. \ref{eq:tauI} and $C_a\sim O(1)$), density wave damping of eccentricity and inclination, and aerodynamic gas drag, as in previous simulations \cite{CanupWard2006}. 
Because the gap-opening mass \cite{crida2006width} exceeds the maximum planet mass (eqn. \ref{eq:Mcrit}), planets are not expected to open gaps and transition to Type-II migration. 

A subset of our simulations consider an inner cavity in the gas disk of radius $r_{\rm cav} = 20R_*$ (in Figure \ref{fig:MassAlpha}) or $r_{\rm cav} = 15R_*$ (in Supplemental discussion \ref{sec:InnerDisk cavity}), such as may result from magnetospheric truncation \cite{zhu2023global} or perhaps from a larger inner disk viscosity \cite{ChatterjeeTan2014}. We do not calculate the detailed changes in the Lindblad and co-rotation torques that occur near the cavity outer edge; these are strongly affected by edge steepness and require modeling of the cavity-producing process \cite{Masset2006,ataiee2021pushing}. We  assume that the cavity properties are such that the co-rotation and Lindblad torques cancel each other once a planet is interior to $r_{\rm cav}$ and deactivate its Type-I migration, similar to previous studies \cite{batygin2023formation,lambrechts2019formation}. \fix{The co-rotation torque responsible for the planet trap \cite{Masset2006} arises from interaction with gas in the local region surrounding the planet's orbit.  The dominant gas interactions responsible for eccentricity and inclination damping arise in this same region \cite{ward1993density}.} We thus assume that planets within the cavity still undergo damping of eccentricity and inclination, as in prior work \cite{batygin2023formation}.  Supplementary Movie 1 and 2 show two example evolutions with and without an inner disk cavity.

Our $N-$body simulations do not account for a gap in the gas disk that can form due to photoevaporative loss (see Figure \ref{fig:DensityOverTime}), which could decrease the migration rate relative to that assumed in the code. Because this gap tends to form late when the total planetary system mass has neared its final value, its presence would not be expected to substantially change the mass ratio values predicted here. 

\subsubsection*{Infall description} 
\textcolor{mod}{In a classic collapse model, the centrifugal radius is a sensitive function of stellar mass, cloud rotation rate ($\omega_{\rm cd}$) and cloud temperature ($T_{\rm cd}$), with $r_c \propto M_*^3\omega_{\rm cd}^2T_{\rm cd}^4 $ \cite{HuesoGuillot2005}. Observations find $7 \le T_c \le 40$ K and few $\times 10^{-15} \le \omega_c \le 10^{-13} \rm s^{-1}$ \cite{li2015masses}, implying a wide range of $r_c$ values, including those comparable to the radial scale of compact systems.  For example, for a $1M_{\odot}$ [$0.08M_{\odot}$] star, $T_{c} = 15$ K, and $\omega_{c}\sim 5\times10^{-15} {\rm s^{-1}}$ [$\omega_c \sim 5 \times 10^{-14} \rm s^{-1}$], $r_c \sim 1$ au [$r_c \sim 0.1$ au] is predicted. Ref\cite{batygin2023formation} argues for $r_c \sim 10^{-1}$ au due to strong magnetic breaking. $N$-body simulations in the main text adopt $r_c = 0.6$ au for a solar mass star, while those in Supplemental discussion \ref{sec:TrapSystem} use $r_c=0.07$ au for an $0.1M_\odot$ star, with these values motivated by the Trappist-1 and Kepler-11 systems.  Similar results to eqns. \ref{eq:Mcrit} through \ref{eq:Mtot_Final} are expected for varied $r_c$ so long as $r_c$ lies beyond the condensation line for rocky materials and planetesimals form interior to $r_c$.}

\textcolor{mod}{Infall is a complex process, and we utilize a simplified, tractable model for the concept development work here. We assume that during infall of the last few percent of the stellar mass to the disk, $r_c$ is constant, the infall flux $F_{\rm in}$ is uniform with radius interior to $r_c$, and that the infall rate decreases with time as $F_{\rm in}(r,t)=F_0\exp(-t/\tau_{\rm in})$.   The total mass delivered during the simulated infall is $M_{\rm in}=\pi r_c^2 F_0 \tau_{\rm in}$, and our accretion simulations consider $M_{\rm in} = 0.03$ to $0.05M_*$ and $\tau_{\rm in} = 5 \times 10^5$  yr.   We find that results depend primarily on the ratio $\tau_g/\tau_{\rm in}$ ($\beta$), rather than on the specific value of $\tau_{\rm in}$, so long as $\tau_{\rm in} \ge {\rm few} \times 10^5$ yr (see above). }

The infall of solid material to the disk interior to $r_c$ is mimicked by the addition of new bodies of mass of about a few $\times 10^{-8}M_*$ to the $N$-body simulation with a rate and position set by $F_{\rm 0}(\varepsilon/f)$, where $f$ is the infall gas-to-solids ratio and $\varepsilon$ is the fraction of infalling solids that are incorporated into planets \cite{CanupWard2006}.  This is consistent with our assumption that planetesimals form interior to $r_c$, discussed next. 

\subsection*{Dust-to-planetesimal accretion} \label{sec:PlanetesimalAcc}

\textcolor{mod}{Formation of $>$ km-sized planetesimals is thought to involve: 1) growth of macroscopic particles (pebbles) via grain-grain collisions, 2) mechanism(s) to spatially concentrate pebbles, and 3) gravitational collapse to yield planetesimals. 
\textcolor{mod_ver2}{While we do not model the dust-to-planetesimal stages of growth here, in this section we make the case that planetesimal formation interior to $r_c$ is plausible given current understanding.}} 


\textcolor{mod}{Collisions between orbiting dust grains initially lead to rapid growth \cite[e.g.,][]{birnstiel2010gas}.  A particle’s Stokes number, $\rm{St}$, is the product of its orbital frequency and its “stopping time” due to gas drag, and $\rm{St}$ increases with particle size with a functional dependence that varies with drag regime.} 

\textcolor{mod}{As particles grow, their collision velocities increase due to size-dependent interactions with the gas disk.  The Stokes number of the largest particles produced via grain collisions, $\rm{St_{max}}$, can be estimated by equating the collision velocity to a critical velocity, $v_f$, above which collisions lead to fragmentation rather than to growth.  With relative velocities dominated by turbulent motions of the gas (and assuming that the $\alpha$ viscosity parameter characterizes the diffusion of both gas and solids), this gives \cite[e.g.,][]{birnstiel2023dust}}

\begin{equation}
    {\rm St_{max}} \approx (v_f/c_s)^2/(3\alpha)
\end{equation}
\textcolor{mod}{with gas sound speed $c_s$ and viscosity parameter $\alpha$. $\rm{St_{max}}$ is strongly dependent on $v_f$, which is an uncertain parameter. \textcolor{mod_ver2}{For water ice, multiple works consider an enhancement of $v_f$ either near the condensation line of water \cite{musiolik2019contacts}, or due to the addition of a thin layer of water molecules \cite{pillich2021drifting}. For rocky grains, multiple} works adopt $v_f=1~\rm{m/s}$ [\citealp[e.g.,][]{birnstiel2010gas,morbidelli2022contemporary,marschall2023inflationary}], but there is a wide range of reported values. Silicate dust collision experiments at room temperatures \cite{guttler2010outcome}, granular mechanics simulations \cite{umstatter2020fragmentation}, and observations of a young massive disk too hot for ices \cite{yamamuro2023massive} imply $v_f$ values in the 10 to 30 m/s range.  Further, new experiments suggest that $v_f$ is increased as temperatures approach the rock melting point (which perhaps is not surprising given increased ``stickiness" of ice-rich grains known to occur near the ice condensation line).  A key quantity in estimating the energy needed to break grain contacts is the surface energy, $\gamma_e$, with $v_f\propto \gamma_e^{5/6}$ \cite{wada2013growth}.  Experiments show that as temperature increases from $1000~\rm{K}$ to $1400~\rm{K}$ \cite{pillich2023composition}, the effective surface energy of silicate grain aggregates increases by 2 to 3 orders-of-magnitude, implying $v_f \sim 50$ to $300~\rm{m/s}$ in the $T\gtrsim 1000$ K disks modeled here. Thus hot inner disks may be a favored region for grain growth, and plausible combinations of $v_f$ and $\alpha$ can yield $10^{-2}<{\rm St_{max}}< 10^{-1}$, with}

\textcolor{mod}{\begin{equation}
    {\rm St_{max}} \approx 0.05\left(\frac{v_f}{30~{\rm m/s}}\right)^2\left(\frac{1200~{\rm K}}{T}\right)\left(\frac{0.001}{\alpha}\right)
\end{equation}}
\textcolor{mod}{Assuming the Epstein drag regime, the corresponding maximum particle size would be}
\textcolor{mod}{\begin{equation}
    a_{\rm max} \approx 10\, {\rm cm} \left(\frac{\sigma_g}{10^3\ \rm{g~ cm^{-2}}}\right)\left(\frac{v_f}{30~{\rm m\ s^{-1}}}\right)^2\left(\frac{1200~{\rm K}}{T}\right)\left(\frac{0.001}{\alpha}\right).
\end{equation}}
\textcolor{mod}{For comparison, prior works often adopt $a_{ \rm max} \sim 10$ cm for icy grain growth in the outer disk, due to the increased stickiness of warm ice near the snowline \cite{morbidelli2022contemporary}.  For rocky grains interior to the snowline, works that adopt $v_f = 1$ m/s find $a_{\rm max} \sim$ mm \cite{morbidelli2022contemporary}, while those that adopt $v_f = 10$ m/s predict $a_{\rm max} \gtrsim$ cm \cite{drazkowska2022planet} or ${\rm St_{max}} \gtrsim O(10^{-1})$ \cite {charnoz2019planetesimal} in inner disk regions. The latter work notes that silicate dust may enter into a plastic regime at temperatures $>1000$ K  that renders every collision more dissipative and increases $v_f$, similar to the effect observed in recent experiments of ref \cite{pillich2023composition}.} 

\textcolor{mod}{Planetesimal formation by steaming instability (SI) requires that the volume density ratio of solids-to-gas exceeds unity in the mid-plane; this condition can be alternatively expressed as a minimum solid-to-gas surface density ratio, $Z_{\rm crit}$.  Determining $Z_{\rm crit}$, which depends on $\rm{St}$ (and also on $\alpha$ if the disk is turbulent), requires state-of-the-art hydrodynamical models with sufficient resolution and run time. Recent high-resolution simulations of laminar disks \cite{li2021thresholds}  find $0.005 < Z_{\rm crit} < 0.006$ for $\rm{St} = 10^{-1}$, and $0.01 \le Z_{\rm crit} < 0.02$ for $\rm{St} = 10^{-2}$, implying that for $\rm{St} \ge 10^{-2}$, collapse may occur with stellar metallicities or only modest solid concentration.  Gas turbulence and particle self-gravity increase $Z_{\rm crit}$ for a given $\rm{St}$. Recent simulations including these effects \cite{lim2024streaming} for $\rm{St} = 10^{-1}$ find $Z_{\rm crit} \sim 0.02$ [$\sim 0.06$] for $\alpha = 10^{-4}$ [$\alpha = 10^{-3}$]; with $\rm{St} = 10^{-2}$, the simulations find $Z_{\rm crit} \sim 0.06$ [$\sim 0.2$] for $\alpha = 10^{-4}$ [$\alpha = 10^{-3}$]. These $Z_{\rm crit}$ values should be viewed as upper limits, because SI and collapse tend to become more likely as numerical resolution is increased \cite{lim2024streaming}}.  

\textcolor{mod}{Several mechanisms could enhance $Z$ in the inner disk.  Inward radial drift of particles formed at larger orbital radii increases $Z$ in the inner disk due to its smaller cross-sectional area \cite{YoudinShu2002}; e.g., dust evolution models \cite[][their Figure 8]{birnstiel2010gas} that include infall find this produces $Z \sim 0.02$ in the disk interior to 1 au.  Larger solid enhancements are possible if particles are concentrated by additional effects.  Hydrodynamical simulations show that during infall, vortices due to Rossby wave instability form near $r_c$, trapping particles and enhancing the local dust-to-gas ratio by a factor of approximately $5$ [10] for 1-cm [10-cm] particles \cite{Bae2015}.} 

\textcolor{mod_ver3}{Another proposal \cite{DrazkowskaSzulagyi2018,batygin2023formation} is that planetesimal formation will be favored in regions where the gas radial velocity is outward and advection of particles by the gas can counter their inner radial decay, leading to a force balance for certain particle sizes that may then become concentrated.   In an infall-supplied disk, the gas velocity becomes outward at a location interior to $r_c$ that is a function of the infall angular momentum; for the uniform infall flux adopted here, this occurs for $r\gtrsim 0.85\,r_c$ for $r_d/r_c \gtrsim 10$ \cite{Canup2002}, where $r_d$ is the disk's outer radius, so that $v_g > 0$ may apply to the outer $\sim 20\%$ of the disk area interior to $r_c$.}      

Additionally, several works propose that solids would be preferentially concentrated near the rock evaporation radius, $r_{\rm evap}$, that falls within $r_c$ in our model \cite{morbidelli2020kuiper, izidoro2022planetesimal}.  Interior to $r_{\rm evap}$, the disk’s effective $\alpha$ value may greatly increase due to magnetically-driven turbulence, creating a local pressure maximum that could trap particles and strongly increase $Z$ \cite{ChatterjeeTan2014}.  \fix{Alternatively,} refs. \cite{morbidelli2022contemporary,batygin2023formation} argue that if the gas velocity is outward across $r_{\rm evap}$, the vapor produced by evaporation of inwardly drifting grains would be carried back outward, allowing for recondensation and a preferred region for solid build-up.   This effect would apply to infall-supplied disks if $r_{\rm evap}$ falls within the outer infall region described above.    

\textcolor{mod}{Thus while tremendous progress in understanding planetesimal formation has been made, substantial uncertainties remain.  Multiple concepts identify disk locales at which solid concentration and planetesimal formation may be preferred (see, e.g., review \cite{drazkowska2022planet}), but models that explicitly treat grain growth up to planetesimal formation are lacking, and whether conditions for collapse are actually achieved is unclear, particularly for turbulent disks.} \textcolor{mod_ver2}{The appropriate  value of $v_f$ in different regimes and/or the importance of an additional bouncing barrier \cite{dominik2024bouncing} remain an important topic of debate.}

\textcolor{mod}{However, observations of young disks increasingly provide evidence of early accretion.  Recent developments highlighted above suggest 1) a higher $v_f$ in inner disks than previously assumed; 2) several mechanisms capable of substantially concentrating solids in inner disks; and 3) a widening of the parameter space that allows for clumping by SI even for $\rm{St} \ll 1$, due to higher-resolution, longer time hydrodynamical simulations.  These results suggest planetesimal formation interior to $r_c$ during final infall is plausible.}  


\backmatter

\bmhead{Data availability}
Output and input files of the N-body simulations shown in Figure \ref{fig:Planetary Systems}a are provided at: https://doi.org/10.5281/zenodo.15232610.The datasets generated during and/or analyzed during the current study are available from the corresponding author upon request. Source data are provided with this paper.

\bmhead{Code availability}

SyMBA is part of the Swift package, which is publicly available at: https://www.boulder.swri.edu/$\sim$hal/swift.html. The modified version of SyMBA utilized here was developed under SwRI internal funds and is thus proprietary.  However, the modifications in this code relative to the standard, freely available SyMBA are described in detail in the Supplementary Information section of Canup \% Ward (2006).


\bmhead{Acknowledgments}
RR gratefully acknowledges the support from a NASA Hubble Fellowship grant (\#HST-HF2-51491).  RMC acknowledges the financial support of NASA's Exoplanets Research Program (80NSSC22K0417). We thank the anonymous reviewers for their comments that improved the final version of this manuscript.

\bmhead{Authors' contributions}

RR and RMC jointly conceived of the paper and carried out the numerical and semi-analytic calculations and their analysis. RR developed the semi-analytic model of gas evolution. RMC carried out the N-body simulations. Both authors contributed to the discussion of the results and to the writing of the manuscript.

\bmhead{Competing interests}

The authors declare no competing interests.

\bmhead{Supplementary information}

The article has accompanying supplementary files. 
\newpage



\section*{\centering Supplemental Material - Origin of compact exoplanetary systems}

\subsubsection*{\centering Raluca Rufu and Robin M. Canup} 
\vspace{2 cm}
{\centering Solar System Science and Exploration Division, Southwest Research Institute, 1301 Walnut St \#400, Boulder, 80302, CO,USA }
\newpage

\begin{appendices}


\setcounter{section}{0}
\begin{suppfigure}[H]
    \centering
    \includegraphics[width=0.5\textwidth]{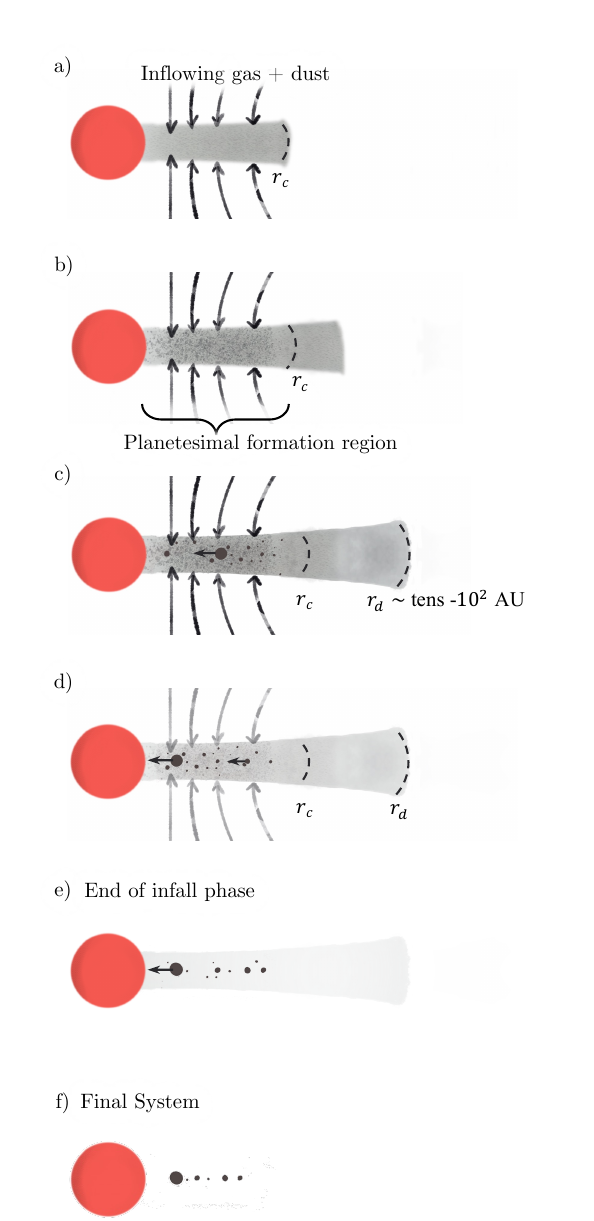}
    
    \caption{\textbf{Schematic of compact system formation during infall.} \textbf{a,} Gas and dust flow into orbits within the centrifugal radius, $r_c$. \textbf{b, c,} Gas viscously spreads inward onto the star and outward past $r_c$, to distances comparable with observed protostellar disk radii $r_d\sim$ few tens to hundreds of au, carrying with it small grains, while larger particles/pebbles decouple from the gas and remain preferentially interior to $r_c$.  Metallicity in the $r\leq r_c$ region increases relative to the infalling material (dark gray region), and planetesimals form via concentration of pebble-sized particles. \textbf{d,} Planetesimals collide and form planets.  As planets grow larger, they migrate inward faster (Type I), and some are lost. Infall resupplies the disk and new planetesimals and planets form, potentially creating multiple generations of planetary systems. \textbf{e,} Infall ends after  $\gtrsim$ few $\times 10^5$ yr \cite{dunham2014evolution}, and the gas disk subsequently disperses via viscous spreading and photoevaporation. Gradually planet migration and growth end. \textbf{f,} Final planetary system has a preferred mass that is $\sim$ few $\times 10^{-5}$ to $10^{-4}$ times the stellar mass, reflecting the balance between infall-regulated accretion and inward disk-driven planet migration. Similar overall results are seen if the gas disk has an inner cavity that stalls migration for planets that orbit within it (see Methods and Supplemental discussion \ref{sec:InnerDisk cavity}).}
    \label{fig:Schematic}
\end{suppfigure}

\clearpage
\section{\mbox{}}\label{sec:DetectabilitySup}
\section*{Detectability threshold}
The signal-to-noise ratio of an observed exoplanet is $S/N\propto P^{-1/3}(R/R_*)^{2}$ (where $P$ is the orbital period of the planet, $R$ is the planetary radius, and $R_*$ is the stellar mass). To approximate  the observational biases and project them onto a simulated population, we define a detectability threshold. We calculate the minimum radius of a planet that is detectable at the simulated orbit \cite{millholland2022edge}:
 \begin{equation}
     \frac{R_{\rm min}}{R_{*,\rm{sim}}}=P_{\rm sim}^{1/6}\times\sqrt{\frac{(S/N)_{\rm min}}{(S/N)_{\rm typical}}}	\left(\frac{R_ p}{R_*}\frac{1}{P_p^{1/6}}\right)_{\rm Kepler-37}
 \end{equation}
where $P_{\rm sim}$ is the period of the simulated planet, $R_ p$ and $P_p$ are the radii and orbital periods of the Kepler-37 planets found by transit (which are among the smaller planets compared to their star, resulting in smaller $R_{\rm min}$ values and an overestimation of the possible detectability), $(S/N)_{\rm typical}\sim30$ is the typical signal-to-noise of Kepler planets \cite{zhu2020patterns}, while the detection threshold of the Kepler pipeline has signal-to-noise of $(S/N)_{\rm min}\sim10$.  We define a simulated planet as being potentially detectable if the radius of the simulated planet (assuming a density of $4\,\rm{g/cm^3}$) is $>R_{\rm min}$ and its period is $<300$ days.

\clearpage
\section{}\label{sec:Mtot_final}
\section*{Analytical estimate of the final planetary system mass ratio}
\setcounter{equation}{0}
The timescale of accretion of a $M_p$ planet is usually assumed to be \cite{lissauer1993growth}: 
\begin{equation}
\begin{aligned}
    \tau_{\rm acc,LS} & =\frac{2}{\pi\sqrt{3}}\frac{1}{\Omega}\frac{M_p}{R_p^2}\frac{f_d}{\sigma_g F_g} \\
        &\sim1.4\times 10^2\ {\rm yr}\left(\frac{M_\odot}{M_*}\right)^{1/2}\left(\frac{r}{0.1\ \rm{au}}\right)^{3/2}\left(\frac{M_p}{M_\oplus}\right)\left(\frac{R_\oplus}{R_p}\right)^2\left(\frac{50}{F_g}\right)\left(\frac{4000\ \rm{g/cm^2}}{\sigma_g/f_d}\right) \label{eqn:tau_acc_LS}
\end{aligned}
\end{equation}
where $\Omega$ is the Keplerian frequency, $r$ is the radial distance from the star, $f_d$ is the disk's gas-to-solid ratio, $F_g$ is a gravitational focusing factor and the surface density, $\sigma_g$ used assumes a minimum-mass extrasolar nebula \cite{chiang2013minimum}. The expression above, however, assumes that the total mass of disk solid material is constant, i.e., that there is no ongoing infall as the planets are forming. 

If instead accretion commences during infall, the solid material available for accretion may be delivered on a timescale that is long compared to $\tau_{\rm acc,LS}$.  In this case, the planetary accretion rate is regulated by the rate at which the infall supplies solid material.  The timescale for the infall to provide a mass $M_p$ in solids to a $2\pi r \Delta r$ area of the disk is:
\begin{equation}
    \tau_{\rm in, solids}=\frac{M_p f}{2\pi r \Delta r F_{\rm in}}
\end{equation}
where $f$ is the gas to solids ratio in the infalling material, $\Delta r$ is the feeding zone width of the planet, and $F_{\rm in}$ is the infall flux. We consider that the spacing between planets is set by the peak planetary eccentricity, with $\Delta r/r\sim2e$ and $e$ reflecting a balance between density wave damping and gravitational scattering \cite{CanupWard2006}:
\begin{equation}
    \frac{\Delta r}{r}\sim 2 \left(\frac{H}{r}\right)\left(\frac{\sigma_s}{\sigma_g}\right)^{1/4}
    \label{eqn:Deltar}
\end{equation}
where $\sigma_s=M_{p}/(2\pi r\Delta r)$ is the surface density of planets, and $H$ is the gas scale height. We assume a uniform infall per radius, so that $F_{\rm in}=\dot{M}_{\rm disk}/(\pi r_c^2)$ (where $\dot{M}_{\rm disk}$ is the total infall rate), so that 
\begin{equation}
\begin{aligned}
    \tau_{\rm in, solids}&=(4\pi)^{-4/5}M_p^{4/5}f\frac{\pi r_c^2}{\dot{M}_{\rm disk}}r^{-8/5}\left(\frac{H}{r}\right)^{-4/5}\sigma_g^{1/5}\\
    &\sim6.4\times10^5\ {\rm yr} \left(\frac{M_p}{M_\oplus}\right)^{4/5}\left(\frac{f}{100}\right)\left(\frac{r_c}{0.6 {\rm au}}\right)^{2}\left(\frac{10^{-7}M_\odot/{\rm yr}}{\dot{M}_{\rm disk}}\right)\\
    &\ \ \ \left(\frac{0.1 {\rm au}}{r}\right)^{8/5}\left(\frac{0.05}{H/r}\right)^{4/5}\left(\frac{\sigma_g}{10^4 {\rm g/cm^2}}\right)^{1/5}.
\end{aligned} \label{eqn:tau_in_solids}
\end{equation}

where here we take typical values found in our 1D gas disk simulations. As seen from eqn. \ref{eqn:tau_acc_LS} and eqn. \ref{eqn:tau_in_solids}, for compact systems, accretion occurs close to the star, hence as $\tau_{\rm acc,LS}\ll\tau_{\rm in, solids}$, planetary accretion would be set by the rate at which solids are being provided to the disk. Accordingly, we define the accretion time scale as: $\tau_{\rm acc}=\tau_{\rm in,solids}/\varepsilon$, where $\varepsilon$ is the fraction of solid material that is ultimately incorporated into planets. The appropriate value for $\varepsilon$ would reflect the total fraction of infalling solids that: i) directly participates in planetesimal formation via local collapse (which could be a relatively small percentage), and ii) is accreted by planetesimals and larger bodies (which would be a high percentage reflecting efficient accretion once large bodies have formed).   

A planet accreting during infall grows mostly in place until it approaches a mass $M_{\rm crit}$ at which $\tau_{\rm acc}\sim\tau_I$, where $\tau_{\rm in}$ is the inward migration timescale.  The planet then migrates inward and may be lost. However while infall is ongoing, a new mass $M_{\rm crit}$ planet may form in its place on a comparable timescale. Therefore, while the infall persists, the system is regulated to contain a few $M_{\rm crit}$ planets at any given time. 

The predicted total planetary system mass ratio at time $t$ is:

\begin{equation}
    \frac{M_{\rm tot}(t)}{M_*} = \int^{r_c}_{r_{\rm in}}\frac{M_{\rm crit}(t)/{M_{*}}}{\Delta r} \rm{d}r
    \label{eq:Mtot_int}
\end{equation}
where $M_*$ is the mass of the star, $r_{\rm in}$ is an inner loss boundary. The gas surface density is assumed to initially reflect a quasi-steady state between infall and viscous spreading \cite{CanupWard2002}, which subsequently decays as a simple exponential on timescale $\tau_{g}$:
\begin{equation}
    \sigma_g(r,t)\sim0.3\frac{F_0}{\alpha\Omega}\left(\frac{r}{H}\right)^2\left( \frac{r_c}{r}\right)^2\exp\left({-t/\tau_g}\right)
    \label{eq:sigma_gas}
\end{equation}
where $F_0$ is the initial infall flux. 

Substituting the solid and gas surface densities into eqn. \ref{eqn:Deltar} gives
\begin{equation}
    \frac{\Delta r}{r} \approx 1.93 \left(\frac{H}{r}\right)^{6/5}\left(\frac{M_{\rm crit}}{M_*}\right)^{1/5}\left(\frac{\alpha\varepsilon}{f}\right)^{1/5}\left(\frac{M_*}{\pi r_c^2F_0}\frac{f}{\varepsilon}\Omega(r_c)\right)^{1/5}\left(\frac{r_c}{r}\right)^{3/10}\exp\left[\frac{1}{5}\frac{t}{\tau_g}\right]
\end{equation}
where $\alpha$ is the disk viscosity parameter, $\varepsilon$ is the fraction of infalling solids incorporated into planets, and $f$ is the infall gas-to-solids ratio. We substitute the above equation and eqn. \ref{eq:Mcrit} into eqn. \ref{eq:Mtot_int} and assume that $(H/r)$ is approximately constant across the disk ($\gamma_c\ll1$, where $\gamma_c$ is defined such that $(H/r) =  (H/r)_{R_*}(r/R_*)^{\gamma_c}$) to find
\begin{equation}
\begin{aligned}
    \frac{M_{\rm tot}(t)}{M_*}&\approx1.5 \left(\frac{\pi}{C_a}\right)^{4/9}\left(\frac{M_*}{\pi r_c^2F_0}\frac{f}{\varepsilon}\Omega(r_c)\right)^{-1/9}\left(\frac{H}{r}\right)^{10/9}\left(\frac{\alpha\varepsilon}{f}\right)^{1/3} \left[1-\left(\frac{r_{\rm in}}{r_c}\right)^{19/18}\right]\\
    & \ \ \ \exp\left[-\frac{t}{\tau_{\rm in}}\left(\frac{4}{9}-\frac{1}{3\beta}\right)\right]\\
    &\approx1.4 \left(\frac{\pi}{C_a}\right)^{4/9}\left(\frac{M_*}{\pi r_c^2F_0}\frac{f}{\varepsilon}\Omega(r_c)\right)^{-1/9}\left(\frac{H}{r}\right)^{10/9}\left(\frac{\alpha\varepsilon}{f}\right)^{1/3}\exp\left[-\frac{t}{\tau_{\rm in}}\left(\frac{4}{9}-\frac{1}{3\beta}\right)\right]
\end{aligned} 
\label{eq:Mtot_apx}
\end{equation}
where $C_a$ is a torque constant, and $\beta\equiv \tau_g/\tau_{\rm in}$, and we set $r_{\rm in}/r_c\sim 0.1$ in the second expression. This is eqn. \ref{eq:Mtot} in the main text. A similar derivation that assumes planets are spaced by a fixed number of Hill radii ($\Delta r\propto r(M_{\rm crit}/3M_*)^{1/3}$) results in $(M_{\rm tot}/M_*) \propto(\alpha\varepsilon/f)^{2/5}e^{-2t/5\tau_{\rm in}(\beta-1)/\beta}$, which has minor differences compared to eqn. \ref{eq:Mtot} for typical parameters used.

As the gas disk surface density and the rate of solid infall decrease, the timescales for migration and accretion both increase. The above-described balance between accretion and migration continues until the migration time for a mass $M_{\rm crit}$ planet is much longer than the gas disk lifetime, i.e., until $\tau_I\sim\tau_{\rm acc}\gg \tau_g$. We find that the transition occurs when the migration timescale of a $M_{\rm crit}$ planet at $r\sim r_c$ is $\tau_I\sim10\tau_g$. The critical mass at the time of this transition is
\begin{equation}
    \eval{\frac{M_{\rm crit}}{M_*}}_{\tau_I=10\tau_g}=\frac{1}{C_a\Omega}\frac{M_*}{r^2\sigma_g(r_c,t)}\left(\frac{H}{r}\right)^2\frac{1}{10\tau_g}
\end{equation}
We substitute eqn. \ref{eq:sigma_gas} into the previous equation and define $M_{\rm in}=\pi r_c^2F_0\tau_{\rm in}$ as the total infalling mass during the final stage of exponentially decreasing infall to find:
\begin{equation}
   \eval{\frac{M_{\rm crit}}{M_*}}_{\tau_I=10\tau_g}\sim\frac{1}{C_a}\left(\frac{\pi\alpha}{0.3}\frac{M_*}{M_{\rm in}}\frac{1}{10\beta}\right) \left(\frac{H}{r}\right)^4\exp\left(t_t/\tau_g\right)
\end{equation}
We substitute the above equation into eqn. \ref{eq:Mcrit} to find the time of transition, $t_t$:
\begin{equation}
\begin{aligned}
    \frac{t_t}{\tau_{\rm in}}&=
     -\frac{9\beta}{5\beta+3}\\
     & \log{\left[0.8\left(\frac{\pi}{C_a}\right)^{4/9}\left(\frac{\alpha\varepsilon}{f}\right)^{1/3}\left(\frac{M_*}{M_{\rm in}}\frac{f/\varepsilon}{10 \beta}\right)\left(\frac{H}{r}\right)^{10/9}\left(\frac{M_*}{M_{\rm in}}\frac{\tau_g}{\beta}\frac{f}{\varepsilon}\Omega(r_c)\right)^{-1/9}\right]}
\end{aligned} \label{eq:t_tauin}
\end{equation}
We set $(M_{\rm tot,0}/M_*)$ as the maximum value from eqn. \ref{eq:Mtot_Final} for $t\ll\tau_{\rm in}$:

\begin{equation}
    \frac{M_{\rm tot,0}}{M_*}\equiv 1.5\left(\frac{\pi}{C_a}\right)^{4/9}\left(\frac{\alpha\varepsilon}{f}\right)^{1/3}\left(\frac{M_*}{M_{\rm in}}\frac{\tau_g}{\beta}\frac{f}{\varepsilon}\Omega(r_c)\right)^{-1/9}\left(\frac{H}{r}\right)^{10/9}\left(1-\left(\frac{r_{\rm in}}{r_c}\right)^{19/18}\right)
\end{equation}

Hence eqn. \ref{eq:t_tauin} may be rewritten as:

\begin{equation}
    \begin{aligned}
    \frac{t_t}{\tau_{\rm in}}&\sim
     -\frac{9\beta}{5\beta+3}\\
      & \log{\left[\frac{1}{2}\frac{M_{\rm tot,0}}{M_*}\frac{1}{1-\left(\frac{r_{\rm in}}{r_c}\right)^{19/18}}\frac{M_*}{M_{\rm in}}\frac{f/\varepsilon}{10 \beta}\right]}
     \end{aligned} 
\end{equation}

Substituting this time into eqn. \ref{eq:Mtot_apx} we get:
\begin{equation}
 \begin{aligned}
     \frac{M_{\rm tot,final}}{M_*} &=  \frac{M_{\rm tot,0}}{M_*} \exp\left(-\frac{t_t}{\tau_{\rm in}}\left(\frac{4}{9}-\frac{1}{3\beta}\right)\right)\\
     & = \left(\frac{M_{\rm tot,0}}{M_*}\right)^{\Lambda+1}\left(\frac{1}{2}\frac{M_*}{M_{\rm in}}\frac{f/\varepsilon}{10 \beta}\frac{1}{1-\left(\frac{r_{\rm in}}{r_c}\right)^{19/18}}\right)^\Lambda\\
     &\sim \left(\frac{M_{\rm tot,0}}{M_*}\right)^{\Lambda+1}\left(\frac{1}{2}\frac{M_*}{M_{\rm in}}\frac{f/\varepsilon}{10 \beta}\right)^\Lambda\\
     \end{aligned} 
 \end{equation}
where $\Lambda\equiv(4\beta-3)/(5\beta+3)$ and we set $r_{\rm in}\ll r_c$ in the last expression.
\clearpage

\section{} \label{sec:SmallerAccRegion}
\section*{Planetesimal accretion in a restricted range of orbital radii}
In the simulations described in the main text, we assume that planetesimals form uniformly throughout the $r<r_c$ region.  However, planetesimal formation may not be uniform across the disk. \textcolor{mod}{In particular, there are several solid concentration mechanisms that could preferentially yield planetesimals in an outer portion of the infall region, including concentration by vorticies that tend to form near $r_c$ \cite{Bae2015} and/or solid concentration near the rock evaporation radius \cite{ChatterjeeTan2014,morbidelli2022contemporary}.  With respect to the latter, we note that while the pebbles necessary for planetesimal formation via SI might only be present at or beyond the rock evaporation radius, large rocky bodies may survive interior to this distance. }  

To assess the potential influence of planetesimal formation in an outer portion of the infall region, we performed additional simulations that assumed that planetesimals accrete in a narrower region, between  $r\sim 0.5$ to $1r_c$, compared to the other simulations. The resulting planetary systems are not significantly different and yield similar mass ratios and planet mass uniformity as the previous set of simulations (Supplementary Figure \ref{fig:InfalRegion}). Although we do not find a notable difference in this set of simulations, further studies that include a time-dependent rock evaporation lines (that determines the inner edge of possible planetesimal accretion), as well as a migration treatment that includes gaps formed after the infall ends (Figure \ref{fig:DensityOverTime}), would be beneficial to better understand the architectures of planetary systems that accrete during infall.

\begin{suppfigure}
    \centering
    \includegraphics[width=1\textwidth]{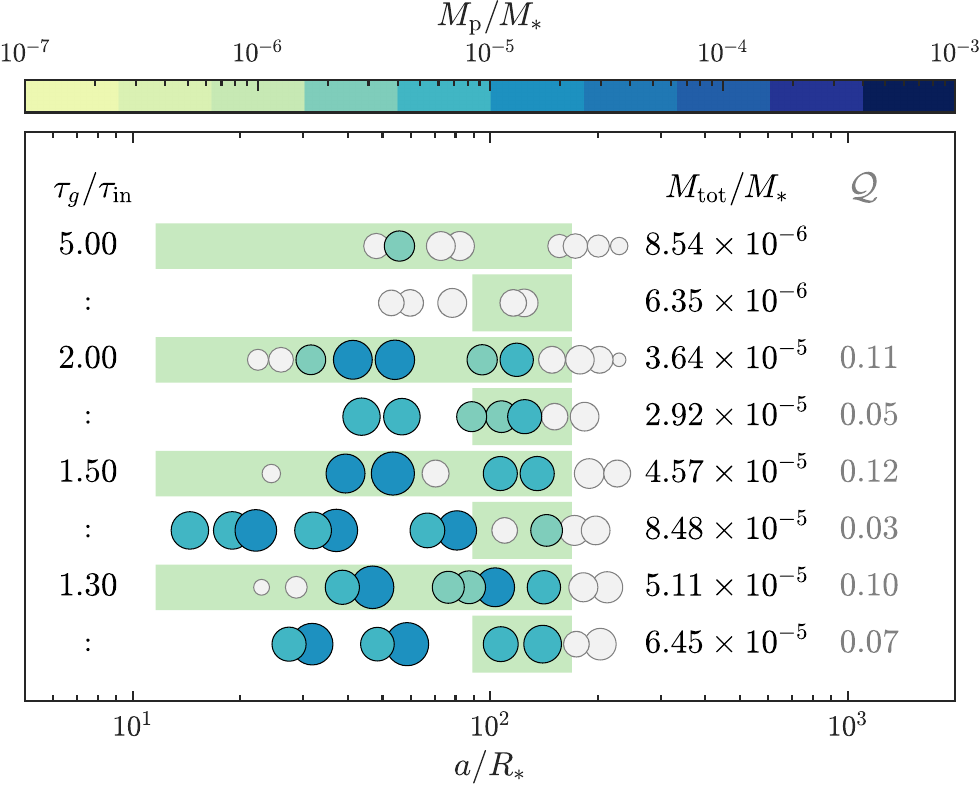}
    \caption{\textbf{Comparison of planetary systems that assume different planetesimal accretion ranges}. Spectral colors represent the planetary mass compared to the stellar mass ($1 M_\odot$), while the grey color represents simulated planets that would not be detectable (see Supplemental Discussion \ref{sec:DetectabilitySup}). The simulations here assume $\alpha\varepsilon/f=5\times10^{-5}$ (where $\alpha$ is the disk viscosity parameter, $\varepsilon$ is the fraction of infalling solids incorporated into planets, and $f$ is the infall gas-to-solids ratio) and different ranges of planetesimal accretion (depicted by the green shaded bars for each simulation): uniform accretion throughout the $r < 0.6$ au region, or a restricted accretion between $0.3$ and $0.6$ au. We merge planets that at the end of the simulation have orbital radii within $3$ mutual Hill radii of each other. The systems are arranged with decreasing $\tau_g/\tau_{\rm in}$ values (listed on the left). The resulting total planetary mass ratio and mass dispersion, $\mathcal{Q}$, are shown on the right side for each planetary system that yield $\geq3$ detectable planets. Source data are provided as a Source Data file.}
    \label{fig:InfalRegion}
\end{suppfigure}
\clearpage

\section{} \label{sec:InnerDisk cavity}
\section*{Accretion simulations with an inner disk cavity}

Magnetohydrodynamics simulations that study the interaction between a disk and a young star indicate that if the star has a sufficiently strong dipole magnetic field, then the gas disk would be depleted close to the star. 

An inner disk cavity (or gas depletion) is suggested to act as a planetary trap for migrating planets \cite{masset2006disk}. As a planet reaches the inner disk edge, the Lindbland and the corotation torque are both increased. If the edge of the disk is narrow enough, the (outward) corotation torque will dominate over the (inward) Lindblad torque right at the disk edge, while an inward total torque would occur farther away, creating a stable fixed point just outside the inner disk edge \cite{masset2006disk}. 

It is unclear whether a cavity with such properties would be present during infall, as Class I stars may instead accrete gas from the disk via boundary layer accretion without an inner cavity \cite{Gaches2024}. 
Further, the extent to which a disk cavity would act as a planetary trap is still uncertain, specifically when multiple planets are present. 
Studies  \cite[e.g.][]{ataiee2021pushing} conclude that the trapping efficiency is heavily dependent on the steepness of the surface density and the planetary masses in the resonant chain. \textcolor{mod_ver2}{In the Trappist-1 system, a cavity may be required to explain the architecture of the system} \cite{huang2022dynamics}. 

While such issues are important to study further, we here considered two simplified limiting cases: 1) full migration, in which all planets undergo inward Type I migration, and 2) disks with an inner cavity that halts Type I migration for any planet that enters the cavity.  
For the latter, we performed simulations with a cavity interior to  $r_{\rm cav}=20R_*$ ($0.13$ au) or $r_{\rm cav}=15R_*$ ($0.10$ au) around a $1M_\odot$ star. We find that the total planetary mass evolution over time (Supplementary Figure \ref{fig:SupMtotVsTime}-a) behaves similarly to the simulations in the main text. Specifically, even if a migration-stopping cavity is included, planets are still lost due to inward migration, because the outer planets continue to grow until they reach a mass capable of driving the inner planet(s) inward (this phenomenon was previously seen also in Ref \cite{Izidoro2017}; their Figure 3). 

The main difference between the two cases is that in the simulations that include a disk cavity, the critical mass threshold marking the onset of planetary loss is somewhat higher compared to the analytical estimate. Hence in the inner disk cavity simulations, the final planetary system mass is somewhat higher than the analytical estimation (eq. \ref{eq:Mtot_Final}) and the previous numerical simulations (Supplementary Figure \ref{fig:SupMtotVsTime}-b). Moreover, the final system mass ratio increases when a wider inner disk cavity is assumed (light red markers compared to red in Supplementary Figure \ref{fig:SupMtotVsTime}-b). 

Supplementary Figure \ref{fig:SupMtotVsAlpha} shows the correlation between the assumed $(\alpha\varepsilon/f)$ and the final system mass ratio given different $\beta$ for cases with a migration-stopping cavity. The final mass ratios are somewhat higher than the estimated values for the full migration case (dashed lines); however, the central dependencies on $(\alpha\varepsilon/f)$ and $\beta \equiv (\tau_g/\tau_{\rm in})$ identified here are still seen. As noted in the main text, the presence of an inner cavity that can stop migration allows a given mass ratio planetary system to result from somewhat less restrictive parameters, e.g., with smaller $\alpha\varepsilon/f$ and/or larger $\beta$. \fix{The resulting system architectures (Supplementary Figure \ref{fig:PlanetarySystems_Cav}) are also similar to the no-cavity cases (Figure \ref{fig:Planetary Systems}a in the main paper)}.
The overall cycle of the planetary accretion and loss, as well as the preference for a common compact system mass ratio independent of stellar mass, are thus seen with or without a migration-stopping inner cavity.

\begin{suppfigure}
      \centering
     \includegraphics[width=0.90\textwidth]{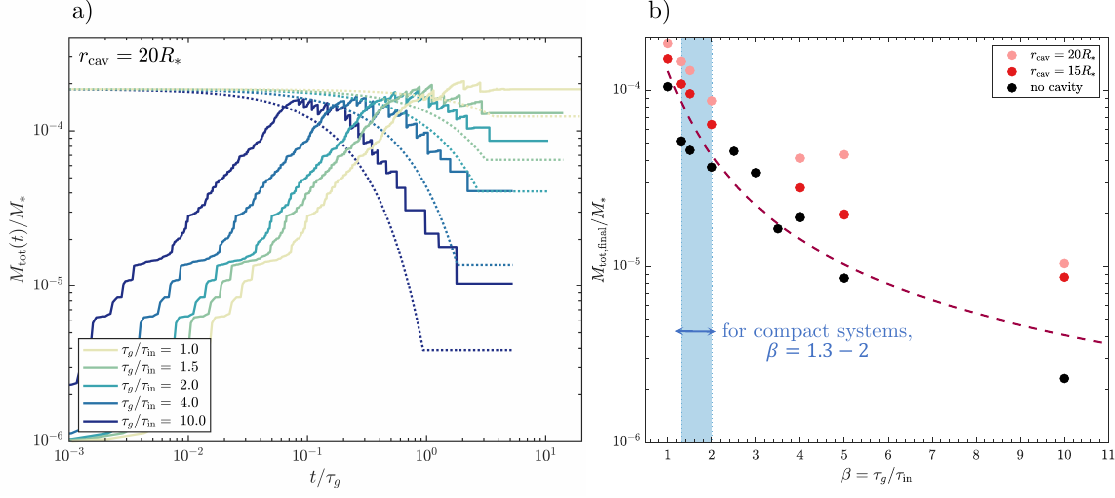}
     \caption{\textcolor{mod}{\textbf{Total planetary system mass scaled to the stellar mass during infall and subsequent gas disk dissipation}. Similar to Figure \ref{fig:MassTime_SameInfall} but includes an inner gas disk cavity. a) Simulations here assume $(\alpha \varepsilon/f)=5\times10^{-5}$ (where $\alpha$ is the disk viscosity parameter, $\varepsilon$ is the fraction of infalling solids incorporated into planets, and $f$ is the infall gas-to-solids ratio) and an inner disk cavity $r_{\rm cav}=20 R_*$. Dotted curves show predictions from eqns. \ref{eq:Mtot} and \ref{eq:Mtot_Final} (note that the analytical solutions do not assume an inner cavity).  \textbf{b,} Final system mass ratio vs. $\beta$  with an inner disk cavity $r_{\rm cav}=20 R_*$ (light red markers), $r_{\rm cav}=15R_*$ (red markers) and no cavity simulations (black markers) compared to analytical prediction (eqn. \ref{eq:Mtot_Final}, dark red dashed line, assuming $f /\varepsilon \sim 10^2$). Source data are provided as a Source Data file.}}\label{fig:SupMtotVsTime}                      
\end{suppfigure}

\begin{suppfigure}
      \centering
     \includegraphics[width=0.65\textwidth]{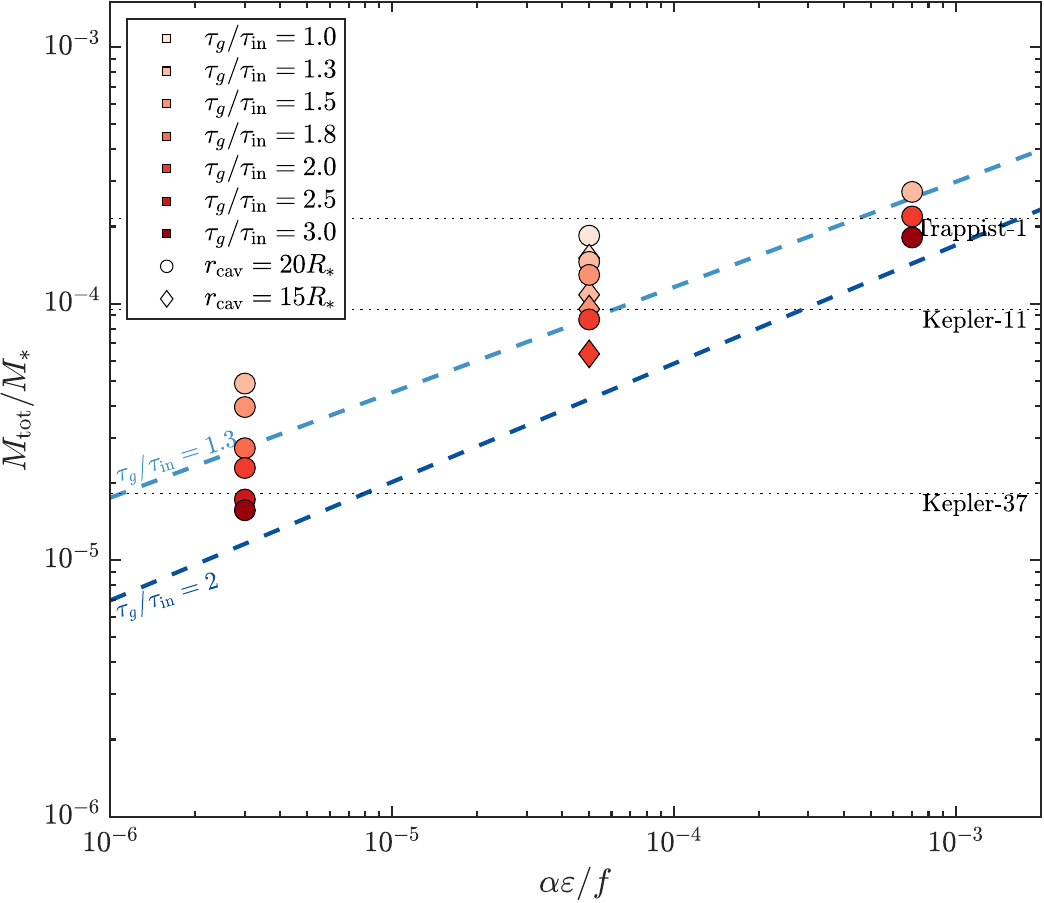}
     \caption{\textbf{Results of planet accretion simulations with different inner disk cavities around a $1M_\odot$ star}. Similar to Figure \ref{fig:MassAlpha} in the main text but with an inner cavity at $r_{\rm cav}=20R_*$ - circles; $r_{\rm cav}=15R_*$ - diamonds. The final total planetary mass is slightly increased compared to the simulations that did not assume an inner disk cavity (Figure \ref{fig:MassAlpha}), however, a similar dependence on $\alpha\varepsilon/f$ is observed. Source data are provided as a Source Data file.} \label{fig:SupMtotVsAlpha} 
 \end{suppfigure}

 \begin{suppfigure}
      \centering
     \includegraphics[width=0.65\textwidth]{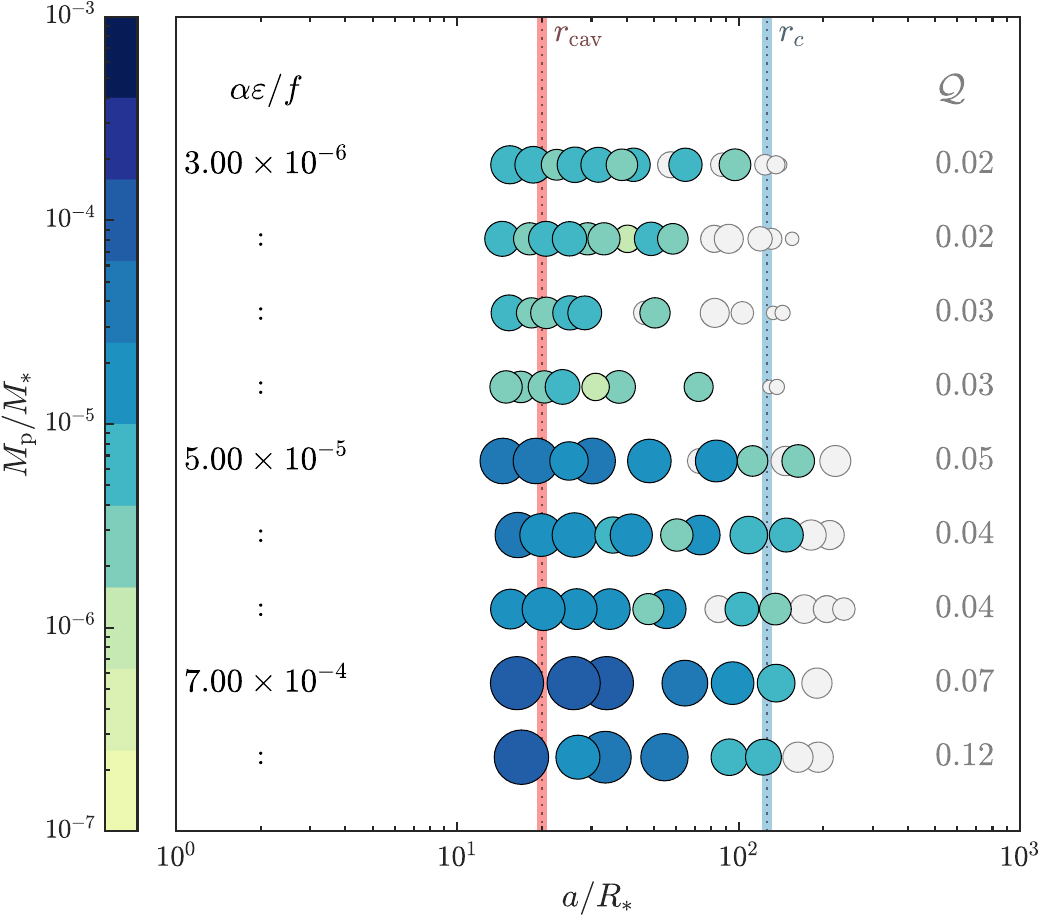}
     \caption{\textcolor{mod}{\textbf{Predicted compact system architectures}. Similar to Figure \ref{fig:Planetary Systems}a 
     in the main text but with an inner cavity at $r_{\rm cav}=20R_*$ (pink line). Systems are arranged by increasing $(\alpha\varepsilon/f)$ and considered varied gas disk lifetimes $1.3\leq\beta\leq2$. Source data are provided as a Source Data file.}} \label{fig:PlanetarySystems_Cav} 
 \end{suppfigure}
 
\clearpage

\section{} \label{sec:TrapSystem}
\section*{Additional simulations}
\begin{suppfigure}[H]
      \centering
    \includegraphics[width=0.75\textwidth]{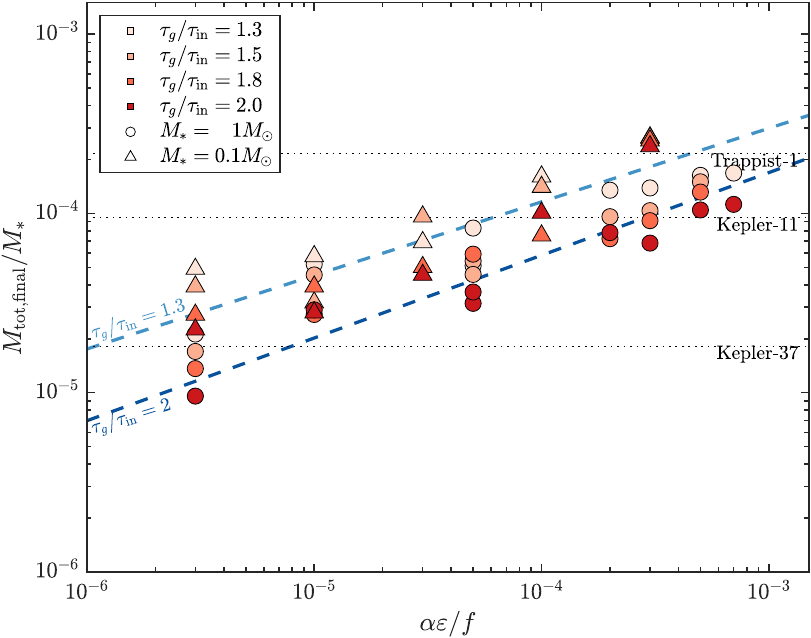}
    \caption{\textbf{Results of planet accretion simulations with varied disk and infall properties}. Similar to Figure \ref{fig:MassAlpha} in the main text but for two stellar masses ($1M_\odot$ - circles; $0.1M_\odot$ - triangles). The infall decays with timescale $\tau_{\rm in}= 5 \times 10^5$ yr, while the gas disk disperses over a longer timescale, $\tau_g=$ 1.3 to 2$\tau_{\rm in}$ (colors, legend). Somewhat larger planetary system masses for a given $(\alpha\varepsilon/f)$ result for the $M_* = 0.1M_{\odot}$ runs due to a larger assumed disk aspect ratio, $H(r_c)/r_c \approx 0.07$, than for the $M_*=1M_{\odot}$ runs that had $H(r_c)/r_c \approx 0.05$ (see eqn. \ref{eq:Mcrit} and Methods). Source data are provided as a Source Data file.} 
     \label{fig:MassAlphaKepTrap}
 \end{suppfigure}

 \begin{suppfigure}
      \centering
     \includegraphics[width=\textwidth]{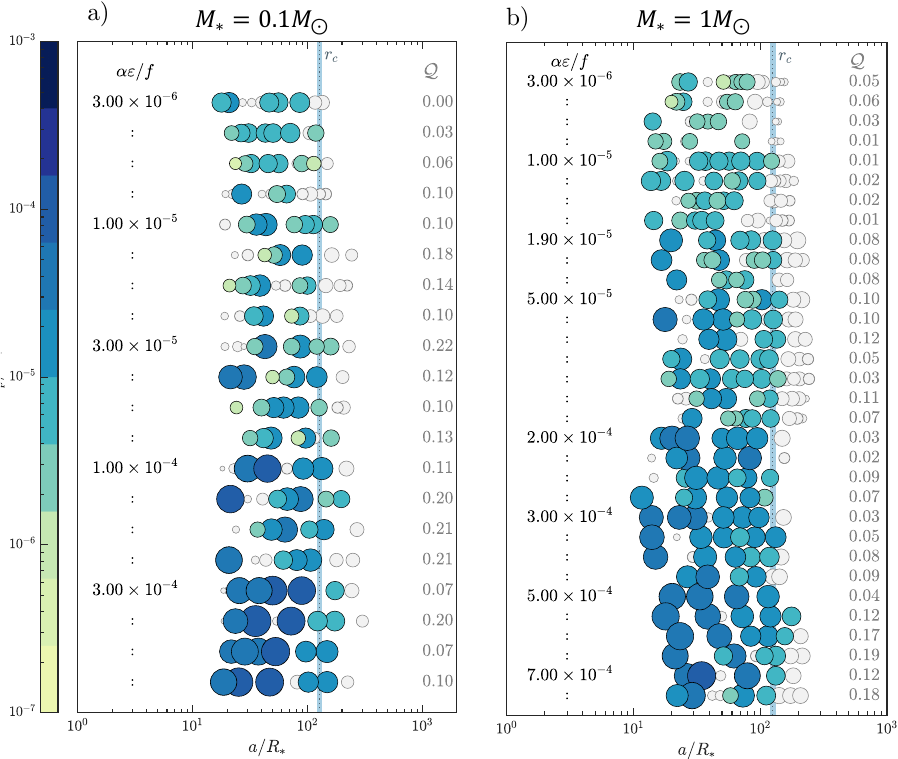}
     \caption{\textbf{Predicted compact system architectures}. Similar to Figure \ref{fig:Planetary Systems}a) in the main text around a) $0.1 M_\odot$ star and b)$1 M_\odot$. Marker sizes are proportional to the planetary mass. Systems are arranged by increasing $(\alpha\varepsilon/f)$ and considered varied gas disk lifetimes ($1.3 \le \beta \leq 2$). Planets with orbital spacings $< 3$ Hill mutual radii that would be unstable on short timescales have been merged; additional collisions may occur on longer timescales. Source data are provided as a Source Data file.\label{fig:Trappist Planetary Systems}} 
\end{suppfigure}
\clearpage

\bibliographystyle{sn-mathphys}

\end{appendices}

\begin{thebibliography}{79}
\ifx \bisbn   \undefined \def \bisbn  #1{ISBN #1}\fi
\ifx \binits  \undefined \def \binits#1{#1}\fi
\ifx \bauthor  \undefined \def \bauthor#1{#1}\fi
\ifx \batitle  \undefined \def \batitle#1{#1}\fi
\ifx \bjtitle  \undefined \def \bjtitle#1{#1}\fi
\ifx \bvolume  \undefined \def \bvolume#1{\textbf{#1}}\fi
\ifx \byear  \undefined \def \byear#1{#1}\fi
\ifx \bissue  \undefined \def \bissue#1{#1}\fi
\ifx \bfpage  \undefined \def \bfpage#1{#1}\fi
\ifx \blpage  \undefined \def \blpage #1{#1}\fi
\ifx \burl  \undefined \def \burl#1{\textsf{#1}}\fi
\ifx \doiurl  \undefined \def \doiurl#1{\url{https://doi.org/#1}}\fi
\ifx \betal  \undefined \def \betal{\textit{et al.}}\fi
\ifx \binstitute  \undefined \def \binstitute#1{#1}\fi
\ifx \binstitutionaled  \undefined \def \binstitutionaled#1{#1}\fi
\ifx \bctitle  \undefined \def \bctitle#1{#1}\fi
\ifx \beditor  \undefined \def \beditor#1{#1}\fi
\ifx \bpublisher  \undefined \def \bpublisher#1{#1}\fi
\ifx \bbtitle  \undefined \def \bbtitle#1{#1}\fi
\ifx \bedition  \undefined \def \bedition#1{#1}\fi
\ifx \bseriesno  \undefined \def \bseriesno#1{#1}\fi
\ifx \blocation  \undefined \def \blocation#1{#1}\fi
\ifx \bsertitle  \undefined \def \bsertitle#1{#1}\fi
\ifx \bsnm \undefined \def \bsnm#1{#1}\fi
\ifx \bsuffix \undefined \def \bsuffix#1{#1}\fi
\ifx \bparticle \undefined \def \bparticle#1{#1}\fi
\ifx \barticle \undefined \def \barticle#1{#1}\fi
\bibcommenthead
\ifx \bconfdate \undefined \def \bconfdate #1{#1}\fi
\ifx \botherref \undefined \def \botherref #1{#1}\fi
\ifx \url \undefined \def \url#1{\textsf{#1}}\fi
\ifx \bchapter \undefined \def \bchapter#1{#1}\fi
\ifx \bbook \undefined \def \bbook#1{#1}\fi
\ifx \bcomment \undefined \def \bcomment#1{#1}\fi
\ifx \oauthor \undefined \def \oauthor#1{#1}\fi
\ifx \citeauthoryear \undefined \def \citeauthoryear#1{#1}\fi
\ifx \endbibitem  \undefined \def \endbibitem {}\fi
\ifx \bconflocation  \undefined \def \bconflocation#1{#1}\fi
\ifx \arxivurl  \undefined \def \arxivurl#1{\textsf{#1}}\fi
\csname PreBibitemsHook\endcsname

\bibitem{weiss2022architectures}
\begin{botherref}
\oauthor{\bsnm{Weiss}, \binits{L.M.}},
\oauthor{\bsnm{Millholland}, \binits{S.C.}},
\oauthor{\bsnm{Petigura}, \binits{E.A.}},
\oauthor{\bsnm{Adams}, \binits{F.C.}},
\oauthor{\bsnm{Batygin}, \binits{K.}},
\oauthor{\bsnm{Bloch}, \binits{A.M.}},
\oauthor{\bsnm{Mordasini}, \binits{C.}}:
Architectures of compact multi-planet systems: Diversity and uniformity.
arXiv preprint arXiv:2203.10076
(2022)
\end{botherref}
\endbibitem

\bibitem{PaardekooperMellema2006}
\begin{barticle}
\bauthor{\bsnm{{Paardekooper}}, \binits{S.-J.}},
\bauthor{\bsnm{{Mellema}}, \binits{G.}}:
\batitle{{Halting {T}ype I planet migration in non-isothermal disks}}.
\bjtitle{\aap}
\bvolume{459}(\bissue{1}),
\bfpage{17}--\blpage{20}
(\byear{2006})
{\href{https://arxiv.org/abs/astro-ph/0608658}{{arXiv:astro-ph/0608658}}}
{[astro-ph]}
\end{barticle}
\endbibitem

\bibitem{OgiharaIda2009}
\begin{barticle}
\bauthor{\bsnm{{Ogihara}}, \binits{M.}},
\bauthor{\bsnm{{Ida}}, \binits{S.}}:
\batitle{N-body simulations of planetary accretion around {M} dwarf stars}.
\bjtitle{\apj}
\bvolume{699}(\bissue{1}),
\bfpage{824}--\blpage{838}
(\byear{2009})
{\href{https://arxiv.org/abs/0904.4543}{{arXiv:0904.4543}}}
{[astro-ph.EP]}
\end{barticle}
\endbibitem

\bibitem{Ormel2017}
\begin{barticle}
\bauthor{\bsnm{{Ormel}}, \binits{C.W.}},
\bauthor{\bsnm{{Liu}}, \binits{B.}},
\bauthor{\bsnm{{Schoonenberg}}, \binits{D.}}:
\batitle{Formation of {TRAPPIST}-1 and other compact systems}.
\bjtitle{\aap}
\bvolume{604},
\bfpage{1}
(\byear{2017})
{\href{https://arxiv.org/abs/1703.06924}{{arXiv:1703.06924}}}
{[astro-ph.EP]}
\end{barticle}
\endbibitem

\bibitem{Izidoro2017}
\begin{barticle}
\bauthor{\bsnm{{Izidoro}}, \binits{A.}},
\bauthor{\bsnm{{Ogihara}}, \binits{M.}},
\bauthor{\bsnm{{Raymond}}, \binits{S.N.}},
\bauthor{\bsnm{{Morbidelli}}, \binits{A.}},
\bauthor{\bsnm{{Pierens}}, \binits{A.}},
\bauthor{\bsnm{{Bitsch}}, \binits{B.}},
\bauthor{\bsnm{{Cossou}}, \binits{C.}},
\bauthor{\bsnm{{Hersant}}, \binits{F.}}:
\batitle{{Breaking the chains: hot super-{E}arth systems from migration and
  disruption of compact resonant chains}}.
\bjtitle{\mnras}
\bvolume{470}(\bissue{2}),
\bfpage{1750}--\blpage{1770}
(\byear{2017})
{\href{https://arxiv.org/abs/1703.03634}{{arXiv:1703.03634}}}
{[astro-ph.EP]}
\end{barticle}
\endbibitem

\bibitem{Izidoro2021}
\begin{botherref}
\oauthor{\bsnm{Izidoro}, \binits{A.}},
\oauthor{\bsnm{Bitsch}, \binits{B.}},
\oauthor{\bsnm{Raymond}, \binits{S.N.}},
\oauthor{\bsnm{Johansen}, \binits{A.}},
\oauthor{\bsnm{Morbidelli}, \binits{A.}},
\oauthor{\bsnm{Lambrechts}, \binits{M.}},
\oauthor{\bsnm{Jacobson}, \binits{S.A.}}:
Formation of planetary systems by pebble accretion and migration. {H}ot
  super-{E}arth systems from breaking compact resonant chains.
A\&A
(2021)
\end{botherref}
\endbibitem

\bibitem{batygin2023formation}
\begin{botherref}
\oauthor{\bsnm{Batygin}, \binits{K.}},
\oauthor{\bsnm{Morbidelli}, \binits{A.}}:
Formation of rocky super-{E}arths from a narrow ring of planetesimals.
Nature Astronomy,
1--9
(2023)
\end{botherref}
\endbibitem

\bibitem{zhu2023global}
\begin{botherref}
\oauthor{\bsnm{Zhu}, \binits{Z.}},
\oauthor{\bsnm{Stone}, \binits{J.M.}},
\oauthor{\bsnm{Calvet}, \binits{N.}}:
A global 3-d simulation of magnetospheric accretion: I. {M}agnetically
  disrupted discs and surface accretion.
Monthly Notices of the Royal Astronomical Society,
3712
(2023)
\end{botherref}
\endbibitem

\bibitem{petigura2018california}
\begin{barticle}
\bauthor{\bsnm{Petigura}, \binits{E.A.}},
\bauthor{\bsnm{Marcy}, \binits{G.W.}},
\bauthor{\bsnm{Winn}, \binits{J.N.}},
\bauthor{\bsnm{Weiss}, \binits{L.M.}},
\bauthor{\bsnm{Fulton}, \binits{B.J.}},
\bauthor{\bsnm{Howard}, \binits{A.W.}},
\bauthor{\bsnm{Sinukoff}, \binits{E.}},
\bauthor{\bsnm{Isaacson}, \binits{H.}},
\bauthor{\bsnm{Morton}, \binits{T.D.}},
\bauthor{\bsnm{Johnson}, \binits{J.A.}}:
\batitle{The {C}alifornia-{K}epler survey. {IV}. {M}etal-rich stars host a
  greater diversity of planets}.
\bjtitle{The Astronomical Journal}
\bvolume{155}(\bissue{2}),
\bfpage{89}
(\byear{2018})
\end{barticle}
\endbibitem

\bibitem{dunham2014evolution}
\begin{botherref}
\oauthor{\bsnm{Dunham}, \binits{M.M.}},
\oauthor{\bsnm{Stutz}, \binits{A.M.}},
\oauthor{\bsnm{Allen}, \binits{L.E.}},
\oauthor{\bsnm{Evans}, \binits{N.}},
\oauthor{\bsnm{Fischer}, \binits{W.J.}},
\oauthor{\bsnm{Megeath}, \binits{S.T.}},
\oauthor{\bsnm{Myers}, \binits{P.C.}},
\oauthor{\bsnm{Offner}, \binits{S.S.}},
\oauthor{\bsnm{Poteet}, \binits{C.A.}},
\oauthor{\bsnm{Tobin}, \binits{J.J.}}, et al.:
The evolution of protostars: Insights from ten years of infrared surveys with
  {S}pitzer and {H}erschel.
Protostars and Planets VI
\textbf{195}
(2014)
\end{botherref}
\endbibitem

\bibitem{kristensen2018protostellar}
\begin{barticle}
\bauthor{\bsnm{Kristensen}, \binits{L.}},
\bauthor{\bsnm{Dunham}, \binits{M.}}:
\batitle{Protostellar half-life: new methodology and estimates}.
\bjtitle{Astronomy \& Astrophysics}
\bvolume{618},
\bfpage{158}
(\byear{2018})
\end{barticle}
\endbibitem

\bibitem{Mottram2017}
\begin{barticle}
\bauthor{\bsnm{{Mottram}}, \binits{J.C.}},
\bauthor{\bsnm{{van Dishoeck}}, \binits{E.F.}},
\bauthor{\bsnm{{Kristensen}}, \binits{L.E.}},
\bauthor{\bsnm{{Karska}}, \binits{A.}},
\bauthor{\bsnm{{San Jos{\'e}-Garc{\'\i}a}}, \binits{I.}},
\bauthor{\bsnm{{Khanna}}, \binits{S.}},
\bauthor{\bsnm{{Herczeg}}, \binits{G.J.}},
\bauthor{\bsnm{{Andr{\'e}}}, \binits{P.}},
\bauthor{\bsnm{{Bontemps}}, \binits{S.}},
\bauthor{\bsnm{{Cabrit}}, \binits{S.}},
\bauthor{\bsnm{{Carney}}, \binits{M.T.}},
\bauthor{\bsnm{{Drozdovskaya}}, \binits{M.N.}},
\bauthor{\bsnm{{Dunham}}, \binits{M.M.}},
\bauthor{\bsnm{{Evans}}, \binits{N.J.}},
\bauthor{\bsnm{{Fedele}}, \binits{D.}},
\bauthor{\bsnm{{Green}}, \binits{J.D.}},
\bauthor{\bsnm{{Harsono}}, \binits{D.}},
\bauthor{\bsnm{{Johnstone}}, \binits{D.}},
\bauthor{\bsnm{{J{\o}rgensen}}, \binits{J.K.}},
\bauthor{\bsnm{{K{\"o}nyves}}, \binits{V.}},
\bauthor{\bsnm{{Nisini}}, \binits{B.}},
\bauthor{\bsnm{{Persson}}, \binits{M.V.}},
\bauthor{\bsnm{{Tafalla}}, \binits{M.}},
\bauthor{\bsnm{{Visser}}, \binits{R.}},
\bauthor{\bsnm{{Y{\i}ld{\i}z}}, \binits{U.A.}}:
\batitle{{Outflows, infall and evolution of a sample of embedded low-mass
  protostars. The {W}illiam {H}erschel {L}ine {L}egacy (WILL) survey}}.
\bjtitle{A\&A}
\bvolume{600},
\bfpage{99}
(\byear{2017})
{\href{https://arxiv.org/abs/1701.04647}{{arXiv:1701.04647}}}
{[astro-ph.SR]}
\end{barticle}
\endbibitem

\bibitem{Manara2018}
\begin{barticle}
\bauthor{\bsnm{{Manara}}, \binits{C.F.}},
\bauthor{\bsnm{{Morbidelli}}, \binits{A.}},
\bauthor{\bsnm{{Guillot}}, \binits{T.}}:
\batitle{{Why do protoplanetary disks appear not massive enough to form the
  known exoplanet population?}}
\bjtitle{\aap}
\bvolume{618},
\bfpage{3}
(\byear{2018})
{\href{https://arxiv.org/abs/1809.07374}{{arXiv:1809.07374}}}
{[astro-ph.EP]}
\end{barticle}
\endbibitem

\bibitem{DashMiguel2020}
\begin{barticle}
\bauthor{\bsnm{{Dash}}, \binits{S.}},
\bauthor{\bsnm{{Miguel}}, \binits{Y.}}:
\batitle{{Planet formation and disc mass dependence in a pebble-driven scenario
  for low-mass stars}}.
\bjtitle{\mnras}
\bvolume{499}(\bissue{3}),
\bfpage{3510}--\blpage{3521}
(\byear{2020})
{\href{https://arxiv.org/abs/2009.14228}{{arXiv:2009.14228}}}
{[astro-ph.EP]}
\end{barticle}
\endbibitem

\bibitem{tychoniec2020dust}
\begin{barticle}
\bauthor{\bsnm{Tychoniec}, \binits{{\L}.}},
\bauthor{\bsnm{Manara}, \binits{C.F.}},
\bauthor{\bsnm{Rosotti}, \binits{G.P.}},
\bauthor{\bparticle{van} \bsnm{Dishoeck}, \binits{E.F.}},
\bauthor{\bsnm{Cridland}, \binits{A.J.}},
\bauthor{\bsnm{Hsieh}, \binits{T.-H.}},
\bauthor{\bsnm{Murillo}, \binits{N.M.}},
\bauthor{\bsnm{Segura-Cox}, \binits{D.}},
\bauthor{\bparticle{van} \bsnm{Terwisga}, \binits{S.E.}},
\bauthor{\bsnm{Tobin}, \binits{J.J.}}:
\batitle{Dust masses of young disks: constraining the initial solid reservoir
  for planet formation}.
\bjtitle{Astronomy \& Astrophysics}
\bvolume{640},
\bfpage{19}
(\byear{2020})
\end{barticle}
\endbibitem

\bibitem{Miotello2014}
\begin{barticle}
\bauthor{\bsnm{{Miotello}}, \binits{A.}},
\bauthor{\bsnm{{Testi}}, \binits{L.}},
\bauthor{\bsnm{{Lodato}}, \binits{G.}},
\bauthor{\bsnm{{Ricci}}, \binits{L.}},
\bauthor{\bsnm{{Rosotti}}, \binits{G.}},
\bauthor{\bsnm{{Brooks}}, \binits{K.}},
\bauthor{\bsnm{{Maury}}, \binits{A.}},
\bauthor{\bsnm{{Natta}}, \binits{A.}}:
\batitle{{Grain growth in the envelopes and disks of Class I protostars}}.
\bjtitle{\aap}
\bvolume{567},
\bfpage{32}
(\byear{2014})
{\href{https://arxiv.org/abs/1405.0821}{{arXiv:1405.0821}}}
{[astro-ph.SR]}
\end{barticle}
\endbibitem

\bibitem{Harsono2018}
\begin{barticle}
\bauthor{\bsnm{{Harsono}}, \binits{D.}},
\bauthor{\bsnm{{Bjerkeli}}, \binits{P.}},
\bauthor{\bsnm{{van der Wiel}}, \binits{M.H.D.}},
\bauthor{\bsnm{{Ramsey}}, \binits{J.P.}},
\bauthor{\bsnm{{Maud}}, \binits{L.T.}},
\bauthor{\bsnm{{Kristensen}}, \binits{L.E.}},
\bauthor{\bsnm{{J{\o}rgensen}}, \binits{J.K.}}:
\batitle{{Evidence for the start of planet formation in a young circumstellar
  disk}}.
\bjtitle{Nature Astronomy}
\bvolume{2},
\bfpage{646}--\blpage{651}
(\byear{2018})
{\href{https://arxiv.org/abs/1806.09649}{{arXiv:1806.09649}}}
{[astro-ph.SR]}
\end{barticle}
\endbibitem

\bibitem{SeguraCox2020}
\begin{barticle}
\bauthor{\bsnm{{Segura-Cox}}, \binits{D.M.}},
\bauthor{\bsnm{{Schmiedeke}}, \binits{A.}},
\bauthor{\bsnm{{Pineda}}, \binits{J.E.}},
\bauthor{\bsnm{{Stephens}}, \binits{I.W.}},
\bauthor{\bsnm{{Fern{\'a}ndez-L{\'o}pez}}, \binits{M.}},
\bauthor{\bsnm{{Looney}}, \binits{L.W.}},
\bauthor{\bsnm{{Caselli}}, \binits{P.}},
\bauthor{\bsnm{{Li}}, \binits{Z.-Y.}},
\bauthor{\bsnm{{Mundy}}, \binits{L.G.}},
\bauthor{\bsnm{{Kwon}}, \binits{W.}},
\bauthor{\bsnm{{Harris}}, \binits{R.J.}}:
\batitle{{Four annular structures in a protostellar disk less than 500,000
  years old}}.
\bjtitle{\nat}
\bvolume{586}(\bissue{7828}),
\bfpage{228}--\blpage{231}
(\byear{2020})
{\href{https://arxiv.org/abs/2010.03657}{{arXiv:2010.03657}}}
{[astro-ph.EP]}
\end{barticle}
\endbibitem

\bibitem{Alves2020}
\begin{barticle}
\bauthor{\bsnm{Alves}, \binits{F.O.}},
\bauthor{\bsnm{Cleeves}, \binits{L.I.}},
\bauthor{\bsnm{Girart}, \binits{J.M.}},
\bauthor{\bsnm{Zhu}, \binits{Z.}},
\bauthor{\bsnm{Franco}, \binits{G.A.}},
\bauthor{\bsnm{Zurlo}, \binits{A.}},
\bauthor{\bsnm{Caselli}, \binits{P.}}:
\batitle{A case of simultaneous star and planet formation}.
\bjtitle{Astrophys. J. Lett}
\bvolume{904}(\bissue{1}),
\bfpage{6}
(\byear{2020})
\end{barticle}
\endbibitem

\bibitem{HuesoGuillot2005}
\begin{barticle}
\bauthor{\bsnm{{Hueso}}, \binits{R.}},
\bauthor{\bsnm{{Guillot}}, \binits{T.}}:
\batitle{Evolution of protoplanetary disks: constraints from {DM} {T}auri and
  {GM} {A}urigae}.
\bjtitle{\aap}
\bvolume{442}(\bissue{2}),
\bfpage{703}--\blpage{725}
(\byear{2005})
{\href{https://arxiv.org/abs/astro-ph/0506496}{{arXiv:astro-ph/0506496}}}
{[astro-ph]}
\end{barticle}
\endbibitem

\bibitem{Hennebelle2016}
\begin{barticle}
\bauthor{\bsnm{Hennebelle}, \binits{P.}},
\bauthor{\bsnm{Commer{\c{c}}on}, \binits{B.}},
\bauthor{\bsnm{Chabrier}, \binits{G.}},
\bauthor{\bsnm{Marchand}, \binits{P.}}:
\batitle{Magnetically self-regulated formation of early protoplanetary disks}.
\bjtitle{ApJL}
\bvolume{830}(\bissue{1}),
\bfpage{8}
(\byear{2016})
\end{barticle}
\endbibitem

\bibitem{Tobin2020}
\begin{barticle}
\bauthor{\bsnm{Tobin}, \binits{J.J.}},
\bauthor{\bsnm{Sheehan}, \binits{P.D.}},
\bauthor{\bsnm{Megeath}, \binits{S.T.}},
\bauthor{\bsnm{D{\'\i}az-Rodr{\'\i}guez}, \binits{A.K.}},
\bauthor{\bsnm{Offner}, \binits{S.S.}},
\bauthor{\bsnm{Murillo}, \binits{N.M.}},
\bauthor{\bparticle{van't} \bsnm{Hoff}, \binits{M.L.}},
\bauthor{\bsnm{Van~Dishoeck}, \binits{E.F.}},
\bauthor{\bsnm{Osorio}, \binits{M.}},
\bauthor{\bsnm{Anglada}, \binits{G.}}, \betal:
\batitle{The {VLA/ALMA} nascent disk and multiplicity ({VANDAM}) survey of
  orion protostars. {II}. {A} statistical characterization of class 0 and class
  {I} protostellar disks}.
\bjtitle{ApJ}
\bvolume{890}(\bissue{2}),
\bfpage{130}
(\byear{2020})
\end{barticle}
\endbibitem

\bibitem{birnstiel2010gas}
\begin{barticle}
\bauthor{\bsnm{Birnstiel}, \binits{T.}},
\bauthor{\bsnm{Dullemond}, \binits{C.}},
\bauthor{\bsnm{Brauer}, \binits{F.}}:
\batitle{Gas-and dust evolution in protoplanetary disks}.
\bjtitle{A\& A}
\bvolume{513},
\bfpage{79}
(\byear{2010})
\end{barticle}
\endbibitem

\bibitem{birnstiel2023dust}
\begin{botherref}
\oauthor{\bsnm{Birnstiel}, \binits{T.}}:
Dust growth and evolution in protoplanetary disks.
arXiv preprint arXiv:2312.13287
(2023)
\end{botherref}
\endbibitem

\bibitem{pillich2023composition}
\begin{barticle}
\bauthor{\bsnm{Pillich}, \binits{C.}},
\bauthor{\bsnm{Bogdan}, \binits{T.}},
\bauthor{\bsnm{Tasto}, \binits{J.}},
\bauthor{\bsnm{Landers}, \binits{J.}},
\bauthor{\bsnm{Wurm}, \binits{G.}},
\bauthor{\bsnm{Wende}, \binits{H.}}:
\batitle{Composition and sticking of hot chondritic dust in a protoplanetary
  hydrogen atmosphere}.
\bjtitle{The Planetary Science Journal}
\bvolume{4}(\bissue{10}),
\bfpage{195}
(\byear{2023})
\end{barticle}
\endbibitem

\bibitem{liu2021millimeter}
\begin{barticle}
\bauthor{\bsnm{Liu}, \binits{H.B.}},
\bauthor{\bsnm{Tsai}, \binits{A.-L.}},
\bauthor{\bsnm{Chen}, \binits{W.P.}},
\bauthor{\bsnm{Liu}, \binits{J.Z.}},
\bauthor{\bsnm{Zhang}, \binits{X.}},
\bauthor{\bsnm{Ma}, \binits{S.}},
\bauthor{\bsnm{Elbakyan}, \binits{V.}},
\bauthor{\bsnm{Green}, \binits{J.D.}},
\bauthor{\bsnm{Hales}, \binits{A.S.}},
\bauthor{\bsnm{Liu}, \binits{S.-Y.}}, \betal:
\batitle{Millimeter-sized dust grains surviving the water-sublimating
  temperature in the inner 10 au of the {FU} {O}ri disk}.
\bjtitle{The Astrophysical Journal}
\bvolume{923}(\bissue{2}),
\bfpage{270}
(\byear{2021})
\end{barticle}
\endbibitem

\bibitem{yamamoto2014examination}
\begin{barticle}
\bauthor{\bsnm{Yamamoto}, \binits{T.}},
\bauthor{\bsnm{Kadono}, \binits{T.}},
\bauthor{\bsnm{Wada}, \binits{K.}}:
\batitle{An examination of collisional growth of silicate dust in
  protoplanetary disks}.
\bjtitle{The Astrophysical Journal Letters}
\bvolume{783}(\bissue{2}),
\bfpage{36}
(\byear{2014})
\end{barticle}
\endbibitem

\bibitem{zamponi2024exploring}
\begin{barticle}
\bauthor{\bsnm{Zamponi}, \binits{J.}},
\bauthor{\bsnm{Maureira}, \binits{M.J.}},
\bauthor{\bsnm{Liu}, \binits{H.B.}},
\bauthor{\bsnm{Zhao}, \binits{B.}},
\bauthor{\bsnm{Segura-Cox}, \binits{D.}},
\bauthor{\bsnm{Ko}, \binits{C.-L.}},
\bauthor{\bsnm{Caselli}, \binits{P.}}:
\batitle{Exploring the dust grain size and polarization mechanism in the hot
  and massive {C}lass 0 disk {IRAS} 16293-2422 {B}}.
\bjtitle{Astronomy \& Astrophysics}
\bvolume{682},
\bfpage{56}
(\byear{2024})
\end{barticle}
\endbibitem

\bibitem{li2021thresholds}
\begin{barticle}
\bauthor{\bsnm{Li}, \binits{R.}},
\bauthor{\bsnm{Youdin}, \binits{A.N.}}:
\batitle{Thresholds for particle clumping by the streaming instability}.
\bjtitle{The Astrophysical Journal}
\bvolume{919}(\bissue{2}),
\bfpage{107}
(\byear{2021})
\end{barticle}
\endbibitem

\bibitem{lim2024streaming}
\begin{barticle}
\bauthor{\bsnm{Lim}, \binits{J.}},
\bauthor{\bsnm{Simon}, \binits{J.B.}},
\bauthor{\bsnm{Li}, \binits{R.}},
\bauthor{\bsnm{Armitage}, \binits{P.J.}},
\bauthor{\bsnm{Carrera}, \binits{D.}},
\bauthor{\bsnm{Lyra}, \binits{W.}},
\bauthor{\bsnm{Rea}, \binits{D.G.}},
\bauthor{\bsnm{Yang}, \binits{C.-C.}},
\bauthor{\bsnm{Youdin}, \binits{A.N.}}:
\batitle{Streaming instability and turbulence: Conditions for planetesimal
  formation}.
\bjtitle{The Astrophysical Journal}
\bvolume{969}(\bissue{2}),
\bfpage{130}
(\byear{2024})
\end{barticle}
\endbibitem

\bibitem{YoudinShu2002}
\begin{barticle}
\bauthor{\bsnm{{Youdin}}, \binits{A.N.}},
\bauthor{\bsnm{{Shu}}, \binits{F.H.}}:
\batitle{{Planetesimal Formation by Gravitational Instability}}.
\bjtitle{\apj}
\bvolume{580}(\bissue{1}),
\bfpage{494}--\blpage{505}
(\byear{2002})
{\href{https://arxiv.org/abs/astro-ph/0207536}{{arXiv:astro-ph/0207536}}}
{[astro-ph]}
\end{barticle}
\endbibitem

\bibitem{drkazkowska2016close}
\begin{barticle}
\bauthor{\bsnm{{Dr{\k{a}}{\.z}kowska}}, \binits{J.}},
\bauthor{\bsnm{Alibert}, \binits{Y.}},
\bauthor{\bsnm{Moore}, \binits{B.}}:
\batitle{Close-in planetesimal formation by pile-up of drifting pebbles}.
\bjtitle{Astronomy \& Astrophysics}
\bvolume{594},
\bfpage{105}
(\byear{2016})
\end{barticle}
\endbibitem

\bibitem{Bae2015}
\begin{barticle}
\bauthor{\bsnm{{Bae}}, \binits{J.}},
\bauthor{\bsnm{{Hartmann}}, \binits{L.}},
\bauthor{\bsnm{{Zhu}}, \binits{Z.}}:
\batitle{Are protoplanetary disks born with vortices? {R}ossby wave instability
  driven by protostellar infall}.
\bjtitle{\apj}
\bvolume{805}(\bissue{1}),
\bfpage{15}
(\byear{2015})
{\href{https://arxiv.org/abs/1503.02694}{{arXiv:1503.02694}}}
{[astro-ph.EP]}
\end{barticle}
\endbibitem

\bibitem{batygin2020formation}
\begin{barticle}
\bauthor{\bsnm{Batygin}, \binits{K.}},
\bauthor{\bsnm{Morbidelli}, \binits{A.}}:
\batitle{Formation of giant planet satellites}.
\bjtitle{ApJ}
\bvolume{894}(\bissue{2}),
\bfpage{143}
(\byear{2020})
\end{barticle}
\endbibitem

\bibitem{ChatterjeeTan2014}
\begin{barticle}
\bauthor{\bsnm{{Chatterjee}}, \binits{S.}},
\bauthor{\bsnm{{Tan}}, \binits{J.C.}}:
\batitle{{Inside-out Planet Formation}}.
\bjtitle{\apj}
\bvolume{780}(\bissue{1}),
\bfpage{53}
(\byear{2014})
{\href{https://arxiv.org/abs/1306.0576}{{arXiv:1306.0576}}}
{[astro-ph.EP]}
\end{barticle}
\endbibitem

\bibitem{CanupWard2006}
\begin{barticle}
\bauthor{\bsnm{{Canup}}, \binits{R.M.}},
\bauthor{\bsnm{{Ward}}, \binits{W.R.}}:
\batitle{{A common mass scaling for satellite systems of gaseous planets}}.
\bjtitle{\nat}
\bvolume{441}(\bissue{7095}),
\bfpage{834}--\blpage{839}
(\byear{2006})
\end{barticle}
\endbibitem

\bibitem{Cilibrasi2020}
\begin{barticle}
\bauthor{\bsnm{Cilibrasi}, \binits{M.}},
\bauthor{\bsnm{Szul{\'a}gyi}, \binits{J.}},
\bauthor{\bsnm{Grimm}, \binits{S.L.}},
\bauthor{\bsnm{Mayer}, \binits{L.}}:
\batitle{An {N}-body population synthesis framework for the formation of moons
  around {J}upiter-like planets}.
\bjtitle{Monthly Notices of the Royal Astronomical Society}
\bvolume{504}(\bissue{4}),
\bfpage{5455}--\blpage{5474}
(\byear{2021})
\end{barticle}
\endbibitem

\bibitem{CanupWard2002}
\begin{barticle}
\bauthor{\bsnm{{Canup}}, \binits{R.M.}},
\bauthor{\bsnm{{Ward}}, \binits{W.R.}}:
\batitle{{Formation of the Galilean Satellites: Conditions of Accretion}}.
\bjtitle{\aj}
\bvolume{124}(\bissue{6}),
\bfpage{3404}--\blpage{3423}
(\byear{2002})
\end{barticle}
\endbibitem

\bibitem{TanakaWard2004}
\begin{barticle}
\bauthor{\bsnm{{Tanaka}}, \binits{H.}},
\bauthor{\bsnm{{Ward}}, \binits{W.R.}}:
\batitle{Three-dimensional interaction between a planet and an isothermal
  gaseous disk. {II}. {E}ccentricity waves and bending waves}.
\bjtitle{\apj}
\bvolume{602}(\bissue{1}),
\bfpage{388}--\blpage{395}
(\byear{2004})
\end{barticle}
\endbibitem

\bibitem{Duncan1998}
\begin{barticle}
\bauthor{\bsnm{{Duncan}}, \binits{M.J.}},
\bauthor{\bsnm{{Levison}}, \binits{H.F.}},
\bauthor{\bsnm{{Lee}}, \binits{M.H.}}:
\batitle{{A Multiple Time Step Symplectic Algorithm for Integrating Close
  Encounters}}.
\bjtitle{\aj}
\bvolume{116}(\bissue{4}),
\bfpage{2067}--\blpage{2077}
(\byear{1998})
\end{barticle}
\endbibitem

\bibitem{romanova2006magnetospheric}
\begin{barticle}
\bauthor{\bsnm{Romanova}, \binits{M.}},
\bauthor{\bsnm{Lovelace}, \binits{R.}}:
\batitle{The magnetospheric gap and the accumulation of giant planets close to
  a star}.
\bjtitle{ApJ}
\bvolume{645}(\bissue{1}),
\bfpage{73}
(\byear{2006})
\end{barticle}
\endbibitem

\bibitem{Gaches2024}
\begin{barticle}
\bauthor{\bsnm{Gaches}, \binits{B.A.}},
\bauthor{\bsnm{Tan}, \binits{J.C.}},
\bauthor{\bsnm{Rosen}, \binits{A.L.}},
\bauthor{\bsnm{Kuiper}, \binits{R.}}:
\batitle{The high-resolution accretion disks of embedded protostars (hades)
  simulations-{I}. impact of protostellar magnetic fields on accretion modes}.
\bjtitle{Astronomy \& Astrophysics}
\bvolume{692},
\bfpage{219}
(\byear{2024})
\end{barticle}
\endbibitem

\bibitem{schib2021influence}
\begin{barticle}
\bauthor{\bsnm{Schib}, \binits{O.}},
\bauthor{\bsnm{Mordasini}, \binits{C.}},
\bauthor{\bsnm{Wenger}, \binits{N.}},
\bauthor{\bsnm{Marleau}, \binits{G.-D.}},
\bauthor{\bsnm{Helled}, \binits{R.}}:
\batitle{The influence of infall on the properties of protoplanetary
  discs-statistics of masses, sizes, lifetimes, and fragmentation}.
\bjtitle{Astronomy \& Astrophysics}
\bvolume{645},
\bfpage{43}
(\byear{2021})
\end{barticle}
\endbibitem

\bibitem{masset2006disk}
\begin{barticle}
\bauthor{\bsnm{Masset}, \binits{F.}},
\bauthor{\bsnm{Morbidelli}, \binits{A.}},
\bauthor{\bsnm{Crida}, \binits{A.}},
\bauthor{\bsnm{Ferreira}, \binits{J.}}:
\batitle{Disk surface density transitions as protoplanet traps}.
\bjtitle{The Astrophysical Journal}
\bvolume{642}(\bissue{1}),
\bfpage{478}
(\byear{2006})
\end{barticle}
\endbibitem

\bibitem{gilbert2020information}
\begin{barticle}
\bauthor{\bsnm{Gilbert}, \binits{G.J.}},
\bauthor{\bsnm{Fabrycky}, \binits{D.C.}}:
\batitle{An information theoretic framework for classifying exoplanetary system
  architectures}.
\bjtitle{The Astronomical Journal}
\bvolume{159}(\bissue{6}),
\bfpage{281}
(\byear{2020})
\end{barticle}
\endbibitem

\bibitem{WeissMarcy2014}
\begin{barticle}
\bauthor{\bsnm{{Weiss}}, \binits{L.M.}},
\bauthor{\bsnm{{Marcy}}, \binits{G.W.}}:
\batitle{The mass-radius relation for 65 exoplanets smaller than 4 {E}arth
  radii}.
\bjtitle{\apjl}
\bvolume{783}(\bissue{1}),
\bfpage{6}
(\byear{2014})
{\href{https://arxiv.org/abs/1312.0936}{{arXiv:1312.0936}}}
{[astro-ph.EP]}
\end{barticle}
\endbibitem

\bibitem{fabrycky2014architecture}
\begin{barticle}
\bauthor{\bsnm{Fabrycky}, \binits{D.C.}},
\bauthor{\bsnm{Lissauer}, \binits{J.J.}},
\bauthor{\bsnm{Ragozzine}, \binits{D.}},
\bauthor{\bsnm{Rowe}, \binits{J.F.}},
\bauthor{\bsnm{Steffen}, \binits{J.H.}},
\bauthor{\bsnm{Agol}, \binits{E.}},
\bauthor{\bsnm{Barclay}, \binits{T.}},
\bauthor{\bsnm{Batalha}, \binits{N.}},
\bauthor{\bsnm{Borucki}, \binits{W.}},
\bauthor{\bsnm{Ciardi}, \binits{D.R.}}, \betal:
\batitle{Architecture of {K}epler's multi-transiting systems. {II}. new
  investigations with twice as many candidates}.
\bjtitle{The Astrophysical Journal}
\bvolume{790}(\bissue{2}),
\bfpage{146}
(\byear{2014})
\end{barticle}
\endbibitem

\bibitem{matsumoto2020breaking}
\begin{barticle}
\bauthor{\bsnm{Matsumoto}, \binits{Y.}},
\bauthor{\bsnm{Ogihara}, \binits{M.}}:
\batitle{Breaking resonant chains: destabilization of resonant planets due to
  long-term mass evolution}.
\bjtitle{The Astrophysical Journal}
\bvolume{893}(\bissue{1}),
\bfpage{43}
(\byear{2020})
\end{barticle}
\endbibitem

\bibitem{morbidelli2022contemporary}
\begin{barticle}
\bauthor{\bsnm{Morbidelli}, \binits{A.}},
\bauthor{\bsnm{Baillie}, \binits{K.}},
\bauthor{\bsnm{Batygin}, \binits{K.}},
\bauthor{\bsnm{Charnoz}, \binits{S.}},
\bauthor{\bsnm{Guillot}, \binits{T.}},
\bauthor{\bsnm{Rubie}, \binits{D.C.}},
\bauthor{\bsnm{Kleine}, \binits{T.}}:
\batitle{Contemporary formation of early solar system planetesimals at two
  distinct radial locations}.
\bjtitle{Nature Astronomy}
\bvolume{6}(\bissue{1}),
\bfpage{72}--\blpage{79}
(\byear{2022})
\end{barticle}
\endbibitem

\bibitem{LuquePalle2022}
\begin{barticle}
\bauthor{\bsnm{Luque}, \binits{R.}},
\bauthor{\bsnm{Pall{\'e}}, \binits{E.}}:
\batitle{Density, not radius, separates rocky and water-rich small planets
  orbiting m dwarf stars}.
\bjtitle{Science}
\bvolume{377}(\bissue{6611}),
\bfpage{1211}--\blpage{1214}
(\byear{2022})
\end{barticle}
\endbibitem

\bibitem{Owen2012}
\begin{barticle}
\bauthor{\bsnm{{Owen}}, \binits{J.E.}},
\bauthor{\bsnm{{Clarke}}, \binits{C.J.}},
\bauthor{\bsnm{{Ercolano}}, \binits{B.}}:
\batitle{{On the theory of disc photoevaporation}}.
\bjtitle{\mnras}
\bvolume{422}(\bissue{3}),
\bfpage{1880}--\blpage{1901}
(\byear{2012})
{\href{https://arxiv.org/abs/1112.1087}{{arXiv:1112.1087}}}
{[astro-ph.SR]}
\end{barticle}
\endbibitem

\bibitem{preibisch2005evolution}
\begin{barticle}
\bauthor{\bsnm{Preibisch}, \binits{T.}},
\bauthor{\bsnm{Feigelson}, \binits{E.D.}}:
\batitle{The evolution of {X}-ray emission in young stars}.
\bjtitle{The Astrophysical Journal Supplement Series}
\bvolume{160}(\bissue{2}),
\bfpage{390}
(\byear{2005})
\end{barticle}
\endbibitem

\bibitem{nakamoto1994formation}
\begin{barticle}
\bauthor{\bsnm{Nakamoto}, \binits{T.}},
\bauthor{\bsnm{Nakagawa}, \binits{Y.}}:
\batitle{Formation, early evolution, and gravitational stability of
  protoplanetary disks}.
\bjtitle{ApJ}
\bvolume{421},
\bfpage{640}--\blpage{650}
(\byear{1994})
\end{barticle}
\endbibitem

\bibitem{BathPringle1981}
\begin{barticle}
\bauthor{\bsnm{{Bath}}, \binits{G.T.}},
\bauthor{\bsnm{{Pringle}}, \binits{J.E.}}:
\batitle{{The evolution of viscous discs. I - Mass transfer variations}}.
\bjtitle{\mnras}
\bvolume{194},
\bfpage{967}--\blpage{986}
(\byear{1981})
\end{barticle}
\endbibitem

\bibitem{kratter2016gravitational}
\begin{barticle}
\bauthor{\bsnm{Kratter}, \binits{K.}},
\bauthor{\bsnm{Lodato}, \binits{G.}}:
\batitle{Gravitational instabilities in circumstellar disks}.
\bjtitle{Annual Review of Astronomy and Astrophysics}
\bvolume{54},
\bfpage{271}--\blpage{311}
(\byear{2016})
\end{barticle}
\endbibitem

\bibitem{Cossou2014}
\begin{barticle}
\bauthor{\bsnm{{Cossou}}, \binits{C.}},
\bauthor{\bsnm{{Raymond}}, \binits{S.N.}},
\bauthor{\bsnm{{Hersant}}, \binits{F.}},
\bauthor{\bsnm{{Pierens}}, \binits{A.}}:
\batitle{{Hot super-Earths and giant planet cores from different migration
  histories}}.
\bjtitle{\aap}
\bvolume{569},
\bfpage{56}
(\byear{2014})
{\href{https://arxiv.org/abs/1407.6011}{{arXiv:1407.6011}}}
{[astro-ph.EP]}
\end{barticle}
\endbibitem

\bibitem{crida2006width}
\begin{barticle}
\bauthor{\bsnm{Crida}, \binits{A.}},
\bauthor{\bsnm{Morbidelli}, \binits{A.}},
\bauthor{\bsnm{Masset}, \binits{F.}}:
\batitle{On the width and shape of gaps in protoplanetary disks}.
\bjtitle{Icarus}
\bvolume{181}(\bissue{2}),
\bfpage{587}--\blpage{604}
(\byear{2006})
\end{barticle}
\endbibitem

\bibitem{Masset2006}
\begin{barticle}
\bauthor{\bsnm{{Masset}}, \binits{F.S.}},
\bauthor{\bsnm{{D'Angelo}}, \binits{G.}},
\bauthor{\bsnm{{Kley}}, \binits{W.}}:
\batitle{{On the Migration of Protogiant Solid Cores}}.
\bjtitle{\apj}
\bvolume{652}(\bissue{1}),
\bfpage{730}--\blpage{745}
(\byear{2006})
{\href{https://arxiv.org/abs/astro-ph/0607155}{{arXiv:astro-ph/0607155}}}
{[astro-ph]}
\end{barticle}
\endbibitem

\bibitem{ataiee2021pushing}
\begin{barticle}
\bauthor{\bsnm{Ataiee}, \binits{S.}},
\bauthor{\bsnm{Kley}, \binits{W.}}:
\batitle{Pushing planets into an inner cavity by a resonant chain}.
\bjtitle{Astronomy \& Astrophysics}
\bvolume{648},
\bfpage{69}
(\byear{2021})
\end{barticle}
\endbibitem

\bibitem{lambrechts2019formation}
\begin{barticle}
\bauthor{\bsnm{Lambrechts}, \binits{M.}},
\bauthor{\bsnm{Morbidelli}, \binits{A.}},
\bauthor{\bsnm{Jacobson}, \binits{S.A.}},
\bauthor{\bsnm{Johansen}, \binits{A.}},
\bauthor{\bsnm{Bitsch}, \binits{B.}},
\bauthor{\bsnm{Izidoro}, \binits{A.}},
\bauthor{\bsnm{Raymond}, \binits{S.N.}}:
\batitle{Formation of planetary systems by pebble accretion and migration-{H}ow
  the radial pebble flux determines a terrestrial-planet or super-{E}arth
  growth mode}.
\bjtitle{Astronomy \& Astrophysics}
\bvolume{627},
\bfpage{83}
(\byear{2019})
\end{barticle}
\endbibitem

\bibitem{ward1993density}
\begin{barticle}
\bauthor{\bsnm{Ward}, \binits{W.R.}}:
\batitle{Density waves in the solar nebula: planetesimal velocities}.
\bjtitle{Icarus}
\bvolume{106}(\bissue{1}),
\bfpage{274}--\blpage{287}
(\byear{1993})
\end{barticle}
\endbibitem

\bibitem{li2015masses}
\begin{barticle}
\bauthor{\bsnm{Li}, \binits{M.}},
\bauthor{\bsnm{Li}, \binits{X.}}:
\batitle{Masses of discs form from collapse of molecular cloud cores}.
\bjtitle{Mon. Notices Royal Astron. Soc.}
\bvolume{449}(\bissue{3}),
\bfpage{2259}--\blpage{2267}
(\byear{2015})
\end{barticle}
\endbibitem

\bibitem{musiolik2019contacts}
\begin{barticle}
\bauthor{\bsnm{Musiolik}, \binits{G.}},
\bauthor{\bsnm{Wurm}, \binits{G.}}:
\batitle{Contacts of water ice in protoplanetary disks---laboratory
  experiments}.
\bjtitle{The Astrophysical Journal}
\bvolume{873}(\bissue{1}),
\bfpage{58}
(\byear{2019})
\end{barticle}
\endbibitem

\bibitem{pillich2021drifting}
\begin{barticle}
\bauthor{\bsnm{Pillich}, \binits{C.}},
\bauthor{\bsnm{Bogdan}, \binits{T.}},
\bauthor{\bsnm{Landers}, \binits{J.}},
\bauthor{\bsnm{Wurm}, \binits{G.}},
\bauthor{\bsnm{Wende}, \binits{H.}}:
\batitle{Drifting inwards in protoplanetary discs-ii. the effect of water on
  sticking properties at increasing temperatures}.
\bjtitle{Astronomy \& Astrophysics}
\bvolume{652},
\bfpage{106}
(\byear{2021})
\end{barticle}
\endbibitem

\bibitem{marschall2023inflationary}
\begin{barticle}
\bauthor{\bsnm{Marschall}, \binits{R.}},
\bauthor{\bsnm{Morbidelli}, \binits{A.}}:
\batitle{An inflationary disk phase to explain extended protoplanetary dust
  disks}.
\bjtitle{Astronomy \& Astrophysics}
\bvolume{677},
\bfpage{136}
(\byear{2023})
\end{barticle}
\endbibitem

\bibitem{guttler2010outcome}
\begin{barticle}
\bauthor{\bsnm{G{\"u}ttler}, \binits{C.}},
\bauthor{\bsnm{Blum}, \binits{J.}},
\bauthor{\bsnm{Zsom}, \binits{A.}},
\bauthor{\bsnm{Ormel}, \binits{C.W.}},
\bauthor{\bsnm{Dullemond}, \binits{C.P.}}:
\batitle{The outcome of protoplanetary dust growth: pebbles, boulders, or
  planetesimals?-{I}. {M}apping the zoo of laboratory collision experiments}.
\bjtitle{Astronomy \& Astrophysics}
\bvolume{513},
\bfpage{56}
(\byear{2010})
\end{barticle}
\endbibitem

\bibitem{umstatter2020fragmentation}
\begin{barticle}
\bauthor{\bsnm{Umst{\"a}tter}, \binits{P.}},
\bauthor{\bsnm{Urbassek}, \binits{H.M.}}:
\batitle{Fragmentation and energy dissipation in collisions of polydisperse
  granular clusters}.
\bjtitle{Astronomy \& Astrophysics}
\bvolume{633},
\bfpage{24}
(\byear{2020})
\end{barticle}
\endbibitem

\bibitem{yamamuro2023massive}
\begin{barticle}
\bauthor{\bsnm{Yamamuro}, \binits{R.}},
\bauthor{\bsnm{Tanaka}, \binits{K.E.}},
\bauthor{\bsnm{Okuzumi}, \binits{S.}}:
\batitle{Massive protostellar disks as a hot laboratory of silicate grain
  evolution}.
\bjtitle{The Astrophysical Journal}
\bvolume{949}(\bissue{1}),
\bfpage{29}
(\byear{2023})
\end{barticle}
\endbibitem

\bibitem{wada2013growth}
\begin{barticle}
\bauthor{\bsnm{Wada}, \binits{K.}},
\bauthor{\bsnm{Tanaka}, \binits{H.}},
\bauthor{\bsnm{Okuzumi}, \binits{S.}},
\bauthor{\bsnm{Kobayashi}, \binits{H.}},
\bauthor{\bsnm{Suyama}, \binits{T.}},
\bauthor{\bsnm{Kimura}, \binits{H.}},
\bauthor{\bsnm{Yamamoto}, \binits{T.}}:
\batitle{Growth efficiency of dust aggregates through collisions with high mass
  ratios}.
\bjtitle{Astronomy \& Astrophysics}
\bvolume{559},
\bfpage{62}
(\byear{2013})
\end{barticle}
\endbibitem

\bibitem{drazkowska2022planet}
\begin{botherref}
\oauthor{\bsnm{Drazkowska}, \binits{J.}},
\oauthor{\bsnm{Bitsch}, \binits{B.}},
\oauthor{\bsnm{Lambrechts}, \binits{M.}},
\oauthor{\bsnm{Mulders}, \binits{G.D.}},
\oauthor{\bsnm{Harsono}, \binits{D.}},
\oauthor{\bsnm{Vazan}, \binits{A.}},
\oauthor{\bsnm{Liu}, \binits{B.}},
\oauthor{\bsnm{Ormel}, \binits{C.W.}},
\oauthor{\bsnm{Kretke}, \binits{K.}},
\oauthor{\bsnm{Morbidelli}, \binits{A.}}:
Planet formation theory in the era of alma and kepler: from pebbles to
  exoplanets.
arXiv preprint arXiv:2203.09759
(2022)
\end{botherref}
\endbibitem

\bibitem{charnoz2019planetesimal}
\begin{barticle}
\bauthor{\bsnm{Charnoz}, \binits{S.}},
\bauthor{\bsnm{Pignatale}, \binits{F.C.}},
\bauthor{\bsnm{Hyodo}, \binits{R.}},
\bauthor{\bsnm{Mahan}, \binits{B.}},
\bauthor{\bsnm{Chaussidon}, \binits{M.}},
\bauthor{\bsnm{Siebert}, \binits{J.}},
\bauthor{\bsnm{Moynier}, \binits{F.}}:
\batitle{Planetesimal formation in an evolving protoplanetary disk with a dead
  zone}.
\bjtitle{Astronomy \& Astrophysics}
\bvolume{627},
\bfpage{50}
(\byear{2019})
\end{barticle}
\endbibitem

\bibitem{DrazkowskaSzulagyi2018}
\begin{barticle}
\bauthor{\bsnm{{Dr{\k{a}}{\.z}kowska}}, \binits{J.}},
\bauthor{\bsnm{{Szul{\'a}gyi}}, \binits{J.}}:
\batitle{Dust evolution and satellitesimal formation in circumplanetary disks}.
\bjtitle{\apj}
\bvolume{866}(\bissue{2}),
\bfpage{142}
(\byear{2018})
{\href{https://arxiv.org/abs/1807.02638}{{arXiv:1807.02638}}}
{[astro-ph.EP]}
\end{barticle}
\endbibitem

\bibitem{Canup2002}
\begin{barticle}
\bauthor{\bsnm{Canup}, \binits{R.M.}},
\bauthor{\bsnm{Ward}, \binits{W.R.}}:
\batitle{Formation of the {G}alilean satellites: Conditions of accretion}.
\bjtitle{AJ}
\bvolume{124}(\bissue{6}),
\bfpage{3404}
(\byear{2002})
\end{barticle}
\endbibitem

\bibitem{morbidelli2020kuiper}
\begin{botherref}
\oauthor{\bsnm{Morbidelli}, \binits{A.}},
\oauthor{\bsnm{Nesvorn{\`y}}, \binits{D.}}:
Kuiper belt: formation and evolution.
The trans-Neptunian solar system,
25--59
(2020)
\end{botherref}
\endbibitem

\bibitem{izidoro2022planetesimal}
\begin{barticle}
\bauthor{\bsnm{Izidoro}, \binits{A.}},
\bauthor{\bsnm{Dasgupta}, \binits{R.}},
\bauthor{\bsnm{Raymond}, \binits{S.N.}},
\bauthor{\bsnm{Deienno}, \binits{R.}},
\bauthor{\bsnm{Bitsch}, \binits{B.}},
\bauthor{\bsnm{Isella}, \binits{A.}}:
\batitle{Planetesimal rings as the cause of the solar system's planetary
  architecture}.
\bjtitle{Nature Astronomy}
\bvolume{6}(\bissue{3}),
\bfpage{357}--\blpage{366}
(\byear{2022})
\end{barticle}
\endbibitem

\bibitem{dominik2024bouncing}
\begin{barticle}
\bauthor{\bsnm{Dominik}, \binits{C.}},
\bauthor{\bsnm{Dullemond}, \binits{C.}}:
\batitle{The bouncing barrier revisited: Impact on key planet formation
  processes and observational signatures}.
\bjtitle{Astronomy \& Astrophysics}
\bvolume{682},
\bfpage{144}
(\byear{2024})
\end{barticle}
\endbibitem

\bibitem{Hartmann1998}
\begin{barticle}
\bauthor{\bsnm{{Hartmann}}, \binits{L.}},
\bauthor{\bsnm{{Calvet}}, \binits{N.}},
\bauthor{\bsnm{{Gullbring}}, \binits{E.}},
\bauthor{\bsnm{{D'Alessio}}, \binits{P.}}:
\batitle{Accretion and the evolution of {T} tauri disks}.
\bjtitle{\apj}
\bvolume{495}(\bissue{1}),
\bfpage{385}--\blpage{400}
(\byear{1998})
\end{barticle}
\endbibitem

\bibitem{jankovic2019close}
\begin{barticle}
\bauthor{\bsnm{Jankovic}, \binits{M.R.}},
\bauthor{\bsnm{Owen}, \binits{J.E.}},
\bauthor{\bsnm{Mohanty}, \binits{S.}}:
\batitle{Close-in super-{E}arths: The first and the last stages of planet
  formation in an {MRI}-accreting disc}.
\bjtitle{Monthly Notices of the Royal Astronomical Society}
\bvolume{484}(\bissue{2}),
\bfpage{2296}--\blpage{2308}
(\byear{2019})
\end{barticle}
\endbibitem

\bibitem{Lesur2015}
\begin{barticle}
\bauthor{\bsnm{{Lesur}}, \binits{G.}},
\bauthor{\bsnm{{Hennebelle}}, \binits{P.}},
\bauthor{\bsnm{{Fromang}}, \binits{S.}}:
\batitle{{Spiral-driven accretion in protoplanetary discs. I. 2D models}}.
\bjtitle{\aap}
\bvolume{582},
\bfpage{9}
(\byear{2015})
{\href{https://arxiv.org/abs/1509.04859}{{arXiv:1509.04859}}}
{[astro-ph.SR]}
\end{barticle}
\endbibitem

\end{thebibliography}

\begin{thebibliography}{15}
\ifx \bisbn   \undefined \def \bisbn  #1{ISBN #1}\fi
\ifx \binits  \undefined \def \binits#1{#1}\fi
\ifx \bauthor  \undefined \def \bauthor#1{#1}\fi
\ifx \batitle  \undefined \def \batitle#1{#1}\fi
\ifx \bjtitle  \undefined \def \bjtitle#1{#1}\fi
\ifx \bvolume  \undefined \def \bvolume#1{\textbf{#1}}\fi
\ifx \byear  \undefined \def \byear#1{#1}\fi
\ifx \bissue  \undefined \def \bissue#1{#1}\fi
\ifx \bfpage  \undefined \def \bfpage#1{#1}\fi
\ifx \blpage  \undefined \def \blpage #1{#1}\fi
\ifx \burl  \undefined \def \burl#1{\textsf{#1}}\fi
\ifx \doiurl  \undefined \def \doiurl#1{\url{https://doi.org/#1}}\fi
\ifx \betal  \undefined \def \betal{\textit{et al.}}\fi
\ifx \binstitute  \undefined \def \binstitute#1{#1}\fi
\ifx \binstitutionaled  \undefined \def \binstitutionaled#1{#1}\fi
\ifx \bctitle  \undefined \def \bctitle#1{#1}\fi
\ifx \beditor  \undefined \def \beditor#1{#1}\fi
\ifx \bpublisher  \undefined \def \bpublisher#1{#1}\fi
\ifx \bbtitle  \undefined \def \bbtitle#1{#1}\fi
\ifx \bedition  \undefined \def \bedition#1{#1}\fi
\ifx \bseriesno  \undefined \def \bseriesno#1{#1}\fi
\ifx \blocation  \undefined \def \blocation#1{#1}\fi
\ifx \bsertitle  \undefined \def \bsertitle#1{#1}\fi
\ifx \bsnm \undefined \def \bsnm#1{#1}\fi
\ifx \bsuffix \undefined \def \bsuffix#1{#1}\fi
\ifx \bparticle \undefined \def \bparticle#1{#1}\fi
\ifx \barticle \undefined \def \barticle#1{#1}\fi
\bibcommenthead
\ifx \bconfdate \undefined \def \bconfdate #1{#1}\fi
\ifx \botherref \undefined \def \botherref #1{#1}\fi
\ifx \url \undefined \def \url#1{\textsf{#1}}\fi
\ifx \bchapter \undefined \def \bchapter#1{#1}\fi
\ifx \bbook \undefined \def \bbook#1{#1}\fi
\ifx \bcomment \undefined \def \bcomment#1{#1}\fi
\ifx \oauthor \undefined \def \oauthor#1{#1}\fi
\ifx \citeauthoryear \undefined \def \citeauthoryear#1{#1}\fi
\ifx \endbibitem  \undefined \def \endbibitem {}\fi
\ifx \bconflocation  \undefined \def \bconflocation#1{#1}\fi
\ifx \arxivurl  \undefined \def \arxivurl#1{\textsf{#1}}\fi
\csname PreBibitemsHook\endcsname

\bibitem{dunham2014evolution}
\begin{botherref}
\oauthor{\bsnm{Dunham}, \binits{M.M.}},
\oauthor{\bsnm{Stutz}, \binits{A.M.}},
\oauthor{\bsnm{Allen}, \binits{L.E.}},
\oauthor{\bsnm{Evans}, \binits{N.}},
\oauthor{\bsnm{Fischer}, \binits{W.J.}},
\oauthor{\bsnm{Megeath}, \binits{S.T.}},
\oauthor{\bsnm{Myers}, \binits{P.C.}},
\oauthor{\bsnm{Offner}, \binits{S.S.}},
\oauthor{\bsnm{Poteet}, \binits{C.A.}},
\oauthor{\bsnm{Tobin}, \binits{J.J.}}, et al.:
The evolution of protostars: Insights from ten years of infrared surveys with
  {S}pitzer and {H}erschel.
Protostars and Planets VI
\textbf{195}
(2014)
\end{botherref}
\endbibitem

\bibitem{millholland2022edge}
\begin{barticle}
\bauthor{\bsnm{Millholland}, \binits{S.C.}},
\bauthor{\bsnm{He}, \binits{M.Y.}},
\bauthor{\bsnm{Zink}, \binits{J.K.}}:
\batitle{Edge-of-the-multis: evidence for a transition in the outer
  architectures of compact multiplanet systems}.
\bjtitle{The Astronomical Journal}
\bvolume{164}(\bissue{2}),
\bfpage{72}
(\byear{2022})
\end{barticle}
\endbibitem

\bibitem{zhu2020patterns}
\begin{barticle}
\bauthor{\bsnm{Zhu}, \binits{W.}}:
\batitle{On the patterns observed in {K}epler multi-planet systems}.
\bjtitle{AJ}
\bvolume{159}(\bissue{5}),
\bfpage{188}
(\byear{2020})
\end{barticle}
\endbibitem

\bibitem{lissauer1993growth}
\begin{bchapter}
\bauthor{\bsnm{Lissauer}, \binits{J.J.}},
\bauthor{\bsnm{Stewart}, \binits{G.R.}}:
\bctitle{Growth of planets from planetesimals}.
In: \bbtitle{Protostars and Planets III},
pp. \bfpage{1061}--\blpage{1088}
(\byear{1993})
\end{bchapter}
\endbibitem

\bibitem{chiang2013minimum}
\begin{barticle}
\bauthor{\bsnm{Chiang}, \binits{E.}},
\bauthor{\bsnm{Laughlin}, \binits{G.}}:
\batitle{The minimum-mass extrasolar nebula: in situ formation of close-in
  super-earths}.
\bjtitle{Mon. Notices Royal Astron. Soc.}
\bvolume{431}(\bissue{4}),
\bfpage{3444}--\blpage{3455}
(\byear{2013})
\end{barticle}
\endbibitem

\bibitem{CanupWard2006}
\begin{barticle}
\bauthor{\bsnm{{Canup}}, \binits{R.M.}},
\bauthor{\bsnm{{Ward}}, \binits{W.R.}}:
\batitle{{A common mass scaling for satellite systems of gaseous planets}}.
\bjtitle{\nat}
\bvolume{441}(\bissue{7095}),
\bfpage{834}--\blpage{839}
(\byear{2006})
\end{barticle}
\endbibitem

\bibitem{CanupWard2002}
\begin{barticle}
\bauthor{\bsnm{{Canup}}, \binits{R.M.}},
\bauthor{\bsnm{{Ward}}, \binits{W.R.}}:
\batitle{{Formation of the Galilean Satellites: Conditions of Accretion}}.
\bjtitle{\aj}
\bvolume{124}(\bissue{6}),
\bfpage{3404}--\blpage{3423}
(\byear{2002})
\end{barticle}
\endbibitem

\bibitem{Bae2015}
\begin{barticle}
\bauthor{\bsnm{{Bae}}, \binits{J.}},
\bauthor{\bsnm{{Hartmann}}, \binits{L.}},
\bauthor{\bsnm{{Zhu}}, \binits{Z.}}:
\batitle{Are protoplanetary disks born with vortices? {R}ossby wave instability
  driven by protostellar infall}.
\bjtitle{\apj}
\bvolume{805}(\bissue{1}),
\bfpage{15}
(\byear{2015})
{\href{https://arxiv.org/abs/1503.02694}{{arXiv:1503.02694}}}
{[astro-ph.EP]}
\end{barticle}
\endbibitem

\bibitem{ChatterjeeTan2014}
\begin{barticle}
\bauthor{\bsnm{{Chatterjee}}, \binits{S.}},
\bauthor{\bsnm{{Tan}}, \binits{J.C.}}:
\batitle{{Inside-out Planet Formation}}.
\bjtitle{\apj}
\bvolume{780}(\bissue{1}),
\bfpage{53}
(\byear{2014})
{\href{https://arxiv.org/abs/1306.0576}{{arXiv:1306.0576}}}
{[astro-ph.EP]}
\end{barticle}
\endbibitem

\bibitem{morbidelli2022contemporary}
\begin{barticle}
\bauthor{\bsnm{Morbidelli}, \binits{A.}},
\bauthor{\bsnm{Baillie}, \binits{K.}},
\bauthor{\bsnm{Batygin}, \binits{K.}},
\bauthor{\bsnm{Charnoz}, \binits{S.}},
\bauthor{\bsnm{Guillot}, \binits{T.}},
\bauthor{\bsnm{Rubie}, \binits{D.C.}},
\bauthor{\bsnm{Kleine}, \binits{T.}}:
\batitle{Contemporary formation of early solar system planetesimals at two
  distinct radial locations}.
\bjtitle{Nature Astronomy}
\bvolume{6}(\bissue{1}),
\bfpage{72}--\blpage{79}
(\byear{2022})
\end{barticle}
\endbibitem

\bibitem{masset2006disk}
\begin{barticle}
\bauthor{\bsnm{Masset}, \binits{F.}},
\bauthor{\bsnm{Morbidelli}, \binits{A.}},
\bauthor{\bsnm{Crida}, \binits{A.}},
\bauthor{\bsnm{Ferreira}, \binits{J.}}:
\batitle{Disk surface density transitions as protoplanet traps}.
\bjtitle{The Astrophysical Journal}
\bvolume{642}(\bissue{1}),
\bfpage{478}
(\byear{2006})
\end{barticle}
\endbibitem

\bibitem{Gaches2024}
\begin{barticle}
\bauthor{\bsnm{Gaches}, \binits{B.A.}},
\bauthor{\bsnm{Tan}, \binits{J.C.}},
\bauthor{\bsnm{Rosen}, \binits{A.L.}},
\bauthor{\bsnm{Kuiper}, \binits{R.}}:
\batitle{The high-resolution accretion disks of embedded protostars (hades)
  simulations-{I}. impact of protostellar magnetic fields on accretion modes}.
\bjtitle{Astronomy \& Astrophysics}
\bvolume{692},
\bfpage{219}
(\byear{2024})
\end{barticle}
\endbibitem

\bibitem{ataiee2021pushing}
\begin{barticle}
\bauthor{\bsnm{Ataiee}, \binits{S.}},
\bauthor{\bsnm{Kley}, \binits{W.}}:
\batitle{Pushing planets into an inner cavity by a resonant chain}.
\bjtitle{Astronomy \& Astrophysics}
\bvolume{648},
\bfpage{69}
(\byear{2021})
\end{barticle}
\endbibitem

\bibitem{huang2022dynamics}
\begin{barticle}
\bauthor{\bsnm{Huang}, \binits{S.}},
\bauthor{\bsnm{Ormel}, \binits{C.W.}}:
\batitle{The dynamics of the {TRAPPIST}-1 system in the context of its
  formation}.
\bjtitle{Monthly Notices of the Royal Astronomical Society}
\bvolume{511}(\bissue{3}),
\bfpage{3814}--\blpage{3831}
(\byear{2022})
\end{barticle}
\endbibitem

\bibitem{Izidoro2017}
\begin{barticle}
\bauthor{\bsnm{{Izidoro}}, \binits{A.}},
\bauthor{\bsnm{{Ogihara}}, \binits{M.}},
\bauthor{\bsnm{{Raymond}}, \binits{S.N.}},
\bauthor{\bsnm{{Morbidelli}}, \binits{A.}},
\bauthor{\bsnm{{Pierens}}, \binits{A.}},
\bauthor{\bsnm{{Bitsch}}, \binits{B.}},
\bauthor{\bsnm{{Cossou}}, \binits{C.}},
\bauthor{\bsnm{{Hersant}}, \binits{F.}}:
\batitle{{Breaking the chains: hot super-{E}arth systems from migration and
  disruption of compact resonant chains}}.
\bjtitle{\mnras}
\bvolume{470}(\bissue{2}),
\bfpage{1750}--\blpage{1770}
(\byear{2017})
{\href{https://arxiv.org/abs/1703.03634}{{arXiv:1703.03634}}}
{[astro-ph.EP]}
\end{barticle}
\endbibitem

\end{thebibliography}


\end{document}